\documentclass[aps,twocolumn,amsmath,amssymb,floatfix]{revtex4}
\pdfoutput=1
\usepackage{graphicx}
\usepackage{dcolumn}
\usepackage{bm}
\usepackage{amsfonts}
\usepackage{graphicx}
\usepackage{dcolumn}
\usepackage{bm}
\usepackage{amsmath}
\usepackage{amssymb}
\usepackage{epstopdf}
\usepackage[makeroom]{cancel}

\begin{document}
\title{Semiclassical Boltzmann transport theory for multi-Weyl semimetals}
\author{Sanghyun Park$^{1}$}
\author{Seungchan Woo$^{1}$}
\author{E. J. Mele$^2$}
\author{Hongki Min$^{1,2}$}
\email{hmin@snu.ac.kr}
\affiliation{$^1$ Department of Physics and Astronomy, Seoul National University, Seoul 08826, Korea}
\affiliation{$^2$ Department of Physics and Astronomy, University of Pennsylvania, Philadelphia, Pennsylvania 19104, USA}

\date{\today}

\begin{abstract}
Multi-Weyl semimetals (m-WSMs) are a new type of Weyl semimetal that have linear dispersion along one symmetry direction but anisotropic nonlinear dispersion along the two transverse directions with a topological charge larger than one. Using the Boltzmann transport theory and fully incorporating the anisotropy of the system, we study the dc conductivity as a function of carrier density and temperature.  We find that the characteristic density and temperature dependence of the transport coefficients at the level of Boltzmann theory are controlled by the topological charge of the multi-Weyl point and distinguish m-WSMs from their linear Weyl counterparts.
\end{abstract}

\maketitle

{\em Introduction}. 
There has been a growing interest in three-dimensional (3D) analogs of graphene called Weyl semimetals (WSMs) where bands disperse linearly in all directions in momentum space around a twofold point degeneracy.  Most attention has been devoted to novel response functions in elementary WSMs which exhibit a linear dispersion; however, recently it has been realized that these are just the simplest members of a family of multi-Weyl semimetals (m-WSMs) \cite{Xu2011,Fang2012,Huang2016} which are characterized instead by double (triple) Weyl-nodes with a linear dispersion along one symmetry direction but quadratic (cubic) dispersion along the remaining two directions. These multi-Weyl nodes have a topologically protected charge (also referred to as chirality) larger than one, a situation that can be stabilized by point group symmetries \cite{Fang2012}.
 
 Noting that multilayer graphenes with certain stacking patterns support two-dimensional (2D) gapless low energy spectra with high chiralities, these m-WSMs can be regarded as the 3D version of multilayer graphenes.
 One can expect that their modified energy dispersion and spin- or pseudospin-momentum locking textures will have important consequences for various physical properties due both to an enhanced density of states (DOS) and the anisotropy in the energy dispersion, distinguishing m-WSMs from elementary WSMs. In this Rapid Communication, we demonstrate that this emerges already at the level of dc conductivity in the strong scattering limit described by semiclassical Boltzmann transport theory.  The transport properties of conventional linear WSMs have recently been explored theoretically by several authors \cite{Wan2011,Hosur2012,Ryu2014,Ominato2014,Sbierski2014,Skinner2014,Ominato2015,DasSarma2015,Ramakrishnan2015,Roy2016}, and there have been theoretical works on the stability of charge-neutral double-Weyl nodes in the presence of Gaussian disorder \cite{Goswami2015,Bera2016,Sbierski2016} and the thermoelectric transport properties in double-Weyl semimetals \cite{Chen2016}. However, as we show below, the density and temperature dependences of the dc conductivity for m-WSMs require an understanding of the effect of anisotropy in the nonlinear dispersion on the scattering. We develop this theory and find that it predicts characteristic power-law dependences of the conductivity on density and temperature that depend on the topological charge of the Weyl node and distinguish m-WSMs from their linear counterparts.

{\em Model}. 
The low-energy effective Hamiltonian for m-WSMs with chirality $J$ near a single Weyl point is given by \cite{Xu2011,Fang2012,Ahn2016}
\begin{eqnarray}\label{eq:wsm_J}
H_J&=&\varepsilon_0 \left[ \left({k_{-}\over k_0}\right)^J \sigma_{+} + \left({k_{+}\over k_0}\right)^J\sigma_{-}\right]+\hbar v_z k_z \sigma_z,
\end{eqnarray}
where $k_\pm=k_x\pm i k_y$, $\sigma_\pm={1\over 2}\left(\sigma_x\pm i\sigma_y\right)$, $\bm{\sigma}$ are the Pauli matrices acting in the space of the two bands that make contact at the Weyl point,
and $k_0$ and $\varepsilon_0$ are the material-dependent parameters in units of momentum and energy, respectively. For simplicity, here we assumed an axial symmetry around the $k_z$ axis. The eigenenergies of the Hamiltonian are given by $\varepsilon_{\pm}=\pm\varepsilon_0\sqrt{\tilde{k}_{\parallel}^{2J}+c_z^2 \tilde{k}_z^2}$, where $\tilde{\bm{k}}=\bm{k}/k_0$, $\tilde{k}_{\parallel}=\sqrt{\tilde{k}_x^2+\tilde{k}_y^2}$, and $c_z=\hbar v_z k_0/\varepsilon_0$, thus the Hamiltonian $H_J$ has a linear dispersion along the $k_z$ direction for $k_x=k_y=0$, whereas a nonlinear dispersion $\sim k_{\parallel}^J$ along the in-plane direction for $k_z=0$. Note that the system described by the Hamiltonian in Eq.~(\ref{eq:wsm_J}) has a nontrivial topological charge characterized by the chirality index $J$ \cite{Fang2012}. [See Sec.~\ref{sec:eigenstate_dos} in the Supplemental Material (SM) \cite{SM} for the eigenstates and DOS for m-WSMs.]

{\em Boltzmann transport theory in anisotropic systems}. 
We use semiclassical Boltzmann transport theory to calculate the density and temperature dependence of the dc conductivity, which is fundamental in understanding the transport properties of a system. Here we focus on the longitudinal part of the dc conductivity assuming time-reversal symmetry with vanishing Hall conductivities.
The Boltzmann transport theory is known to be valid in the high carrier density limit, and we assume that the Fermi energy is away from the Weyl node, as shown in experiments \cite{Lv2015, Xu2015}. The limitation of the current approach will be discussed later.

For a $d$-dimensional \emph{isotropic} system in which only a single band is involved in the scattering, it is well known that the momentum relaxation time at a wavevector $\bm{k}$ in the relaxation time approximation can be expressed as \cite{Ashcroft1976}
\begin{equation}
\label{eq:relaxation_time_isotropic}
{1\over \tau_{\bm k}}=\int {d^d k' \over (2\pi)^d} W_{\bm{k}\bm{k}'} (1-\cos\theta_{\bm{k}\bm{k}'}),
\end{equation}
where $W_{\bm{k}\bm{k}'}={2\pi\over\hbar} n_{\rm imp} |V_{\bm{k}\bm{k}'}|^2 \delta(\varepsilon_{\bm{k}}-\varepsilon_{\bm{k}'})$, $n_{\rm imp}$ is the impurity density, and $V_{\bm{k}\bm{k}'}$ is the impurity potential describing a scattering from $\bm{k}$ to $\bm{k}'$. The inverse relaxation time is a weighted average of the collision probability in which the forward scattering ($\theta_{\bm{k}\bm{k}'}=0$) receives reduced weight.

For an \emph{anisotropic} system,  the relaxation time approximation Eq.~(\ref{eq:relaxation_time_isotropic}) does not correctly describe the effects of the anisotropy on transport. Instead, coupled integral equations relating  the relaxation times at different angles need to be solved to treat the anisotropy in the nonequilibrium distribution \cite{Schliemann2003,Vyborny2009}. The linearized Boltzmann transport equation for the distribution function $f_{\bm{k}}=f^{(0)}(\varepsilon)+\delta f_{\bm{k}}$ at energy $\varepsilon=\varepsilon_{\bm{k}}$ balances acceleration on the Fermi surface against the scattering rates 
\begin{equation}
(-e)\bm{E}\cdot\bm{v}_{\bm{k}}S^{(0)}(\varepsilon)=\int {d^d k'\over (2\pi)^d} W_{\bm{k}\bm{k}'}\left(\delta f_{\bm{k}}-\delta f_{\bm{k}'}\right),
\end{equation}
where
$S^{(0)}(\varepsilon)=-{\partial f^{(0)}(\varepsilon)\over\partial\varepsilon}$, $f^{(0)}(\varepsilon)=\left[e^{\beta (\varepsilon-\mu)}+ 1\right]^{-1}$ is the Fermi distribution function at equilibrium, and $\beta={1\over k_{\rm B} T}$.
We parametrize $\delta f_{\bm{k}}$ in the form:
\begin{equation}
\delta f_{\bm{k}}=(-e)\left(\sum_{i=1}^{d} E^{(i)} v_{\bm{k}}^{(i)} \tau_{\bm{k}}^{(i)} \right)S^{(0)}(\varepsilon),
\end{equation}
where $E^{(i)}$, $v_{\bm{k}}^{(i)}$, and $\tau_{\bm{k}}^{(i)}$ are the electric field, velocity, and relaxation time along the $i$-th direction, respectively.
After matching each coefficient in $E^{(i)}$, we obtain an integral equation for the relaxation time,
\begin{equation}
\label{eq:relaxation_time_anisotropic}
1=\int {d^d k'\over (2\pi)^d} W_{\bm{k}\bm{k}'}\left(\tau_{\bm{k}}^{(i)}-{v_{\bm{k}'}^{(i)}\over v_{\bm{k}}^{(i)} }\tau_{\bm{k}'}^{(i)}\right).
\end{equation}
For the isotropic case [$\tau_{\bm{k}}^{(i)}=\tau(\varepsilon$) for a given energy $\varepsilon=\varepsilon_{\bm k}$], Eq.~(\ref{eq:relaxation_time_anisotropic}) reduces to Eq.~(\ref{eq:relaxation_time_isotropic}). 
[See Sec.~\ref{sec:conductivity_3D_anisotropic} in SM \cite{SM} for applications of Eq.~(\ref{eq:relaxation_time_anisotropic}) to m-WSMs.]
The current density $\bm{J}$ induced by an electric field $\bm{E}$ is then given by
\begin{equation}
J^{(i)}=g\int {d^d k\over (2\pi)^d} (-e) v_{\bm{k}}^{(i)} \delta f_{\bm{k}}\equiv \sigma_{ij}E^{(j)},
\end{equation}
where $g$ is the degeneracy factor and $\sigma_{ij}$ is the conductivity tensor given by
\begin{equation}
\label{eq:conductivity_tensor}
\sigma_{ij}=g e^2\int {d^d k\over (2\pi)^d} S^{(0)}(\varepsilon) v_{\bm{k}}^{(i)} v_{\bm{k}}^{(j)}\tau_{\bm{k}}^{(j)}.
\end{equation}
For the calculation, we set $g=4$ and $v_z=v_0\equiv {\varepsilon_0 \over \hbar k_0}$.


{\em Density dependence of dc conductivity}. 
Consider the m-WSMs described by Eq.~(\ref{eq:wsm_J}) with chirality $J$ and their dc conductivity as a function of carrier density at zero temperature. Due to the anisotropic energy dispersion with the axial symmetry, for $J>1$ the conductivity also will be anisotropic as $\sigma_{xx}=\sigma_{yy}\neq \sigma_{zz}$. 

 We consider two types of impurity scattering: short-range impurities (e.g., lattice defects, vacancies, and dislocations) and charged impurities distributed randomly in the background. The impurity potential for short-range scatterers is given by a constant $V_{\bm{k}\bm{k}'}=V_{\rm short}$ in momentum space (i.e., zero-range delta function in real space), whereas for charged Coulomb impurities in 3D it is given by
$V_{\bm{k}\bm{k}'}={4\pi e^2 \over \epsilon(\bm{q})|\bm{q}|^2}$,
where $\epsilon(\bm{q})$ is the dielectric function for $\bm{q}=\bm{k}-\bm{k}'$. Within the Thomas-Fermi approximation, the dielectric function can be approximated as $\epsilon(\bm{q})\approx \kappa\left[1+(q_{\rm TF}^2/|\bm{q}|^2)\right]$, where $\kappa$ is the background dielectric constant, $q_{\rm TF}=\sqrt{{4\pi e^2 \over \kappa} D(\varepsilon_{\rm F})}$ is the Thomas-Fermi wave vector, and $D(\varepsilon_{\rm F})$ is the DOS at the Fermi energy $\varepsilon_{\rm F}$. The interaction strength for charged impurities can be characterized by an effective fine structure constant $\alpha={e^2\over \kappa \hbar v_0}$. Note that $q_{\rm TF}\propto\sqrt{g\alpha}$.

\begin{figure}[htb]
\includegraphics[width=1\linewidth]{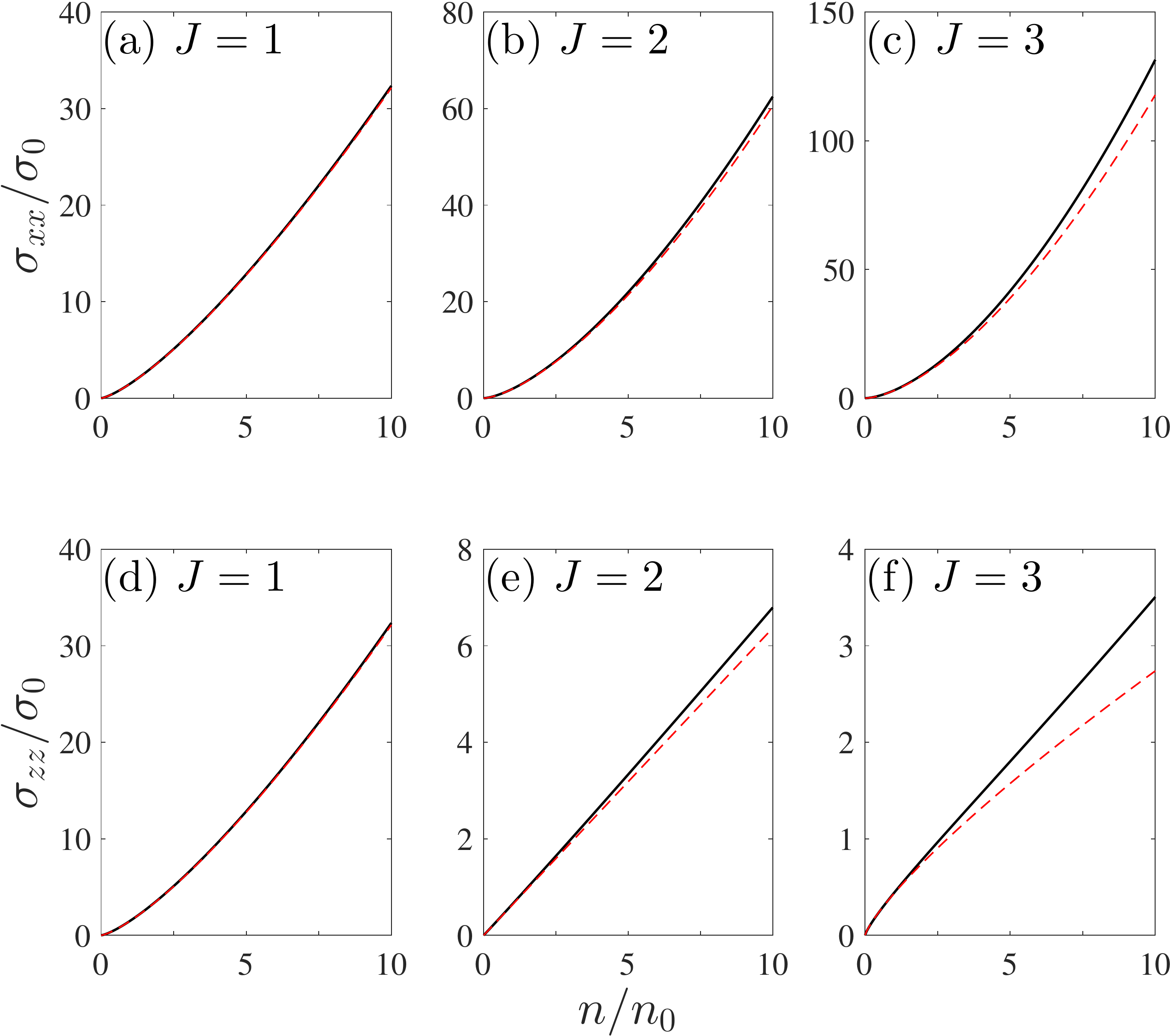}
\caption{
Density dependence of dc conductivity (a)-(c) $\sigma_{xx}$ and (d)-(f) $\sigma_{zz}$ for charged impurities with $g\alpha=1000$. 
Here, $\sigma_0$ and $n_0$ are density-independent normalization constants in units of conductivity and density, respectively, defined in SM \cite{SM}. Red dashed lines represent analytic forms in the strong screening limit given by Eq.~(\ref{eq:sigma_strong_screening_analytic}) in SM \cite{SM}.}
\label{fig:density_dependence_charged}
\end{figure}

Figure \ref{fig:density_dependence_charged} shows the density dependence of the dc conductivity for charged impurity scattering at zero temperature. Because of the chirality $J$, m-WSMs have a characteristic density dependence in dc conductivity, which can be understood as follows. From Eq.~(\ref{eq:conductivity_tensor}), we expect $\sigma_{ii}\sim [v_{\rm F}^{(i)}]^2/V_{\rm F}^2$, where $v_{\rm F}^{(i)}$ is the Fermi velocity along the $i$th direction and $V_{\rm F}^2$ is the angle-averaged squared impurity potential at the Fermi energy $\varepsilon_{\rm F}$.
For m-WSMs, the in-plane component with $k_z=0$ and out-of-plane component with $k_x=k_y=0$ for the velocity at $\varepsilon_{\rm F}$ are given by $v_{\rm F}^{(\parallel)}=J v_0 r_{\rm F}^{1-{1\over J}}$ and $v_{\rm F}^{(z)}=v_0 c_z$, respectively, where $r_{\rm F}=\varepsilon_{\rm F}/\varepsilon_0$. (See Sec.~\ref{sec:eigenstate_dos} in SM \cite{SM}.)

For charged impurities, in the strong screening limit ($g\alpha\gg 1$), $V_{\rm F}\sim q_{\rm TF}^{-2}\sim D^{-1}(\varepsilon_{\rm F})\sim \varepsilon_{\rm F}^{-{2\over J}}$, thus we find
\begin{subequations}\label{eq:density_dependence_charged_strong_screening}
\begin{eqnarray}
\sigma_{xx}&\sim& \varepsilon_{\rm F}^{2\left(1-{1\over J}\right)} \varepsilon_{\rm F}^{4\over J}\sim n^{2(J+1)\over J+2}, \\
\sigma_{zz}&\sim& \varepsilon_{\rm F}^{4\over J}\sim n^{4\over J+2}.
\end{eqnarray}
\end{subequations}
Here, the DOS is $D(\varepsilon)\sim \varepsilon^{2\over J}$, thus $\varepsilon_{\rm F}\sim n^{J\over J+2}$.
In the weak screening limit ($g\alpha\ll 1$), we expect $V_{\rm F}\sim \varepsilon_{\rm F}^{-2\zeta}$ with ${1\over J} \le \zeta \le 1$, because the in-plane and out-of-plane components of the wavevector at $\varepsilon_{\rm F}$ are $k_{\rm F}^{(\parallel)}=k_0 r_{\rm F}^{1\over J}$ and $k_{\rm F}^{(z)}=k_0 r_{\rm F}/c_z$, respectively. Thus, we find
\begin{subequations}\label{eq:density_dependence_charged_weak_screening}
\begin{eqnarray}
\sigma_{xx}&\sim& \varepsilon_{\rm F}^{2\left(1-{1\over J}\right)} \varepsilon_{\rm F}^{4\zeta}\sim n^{2(J-1)+4J\zeta \over J+2}, \\
\sigma_{zz}&\sim& \varepsilon_{\rm F}^{4\zeta}\sim n^{4J\zeta \over J+2}.
\end{eqnarray}
\end{subequations}
(See Sec.~\ref{sec:conductivity_3D_anisotropic} in SM \cite{SM} for the analytic expressions of the dc conductivity for short-range impurities and for charged impurities in the strong screening limit, and a detailed discussion for charged impurities in the weak screening limit.)

\begin{figure}[htb]
\includegraphics[width=1\linewidth]{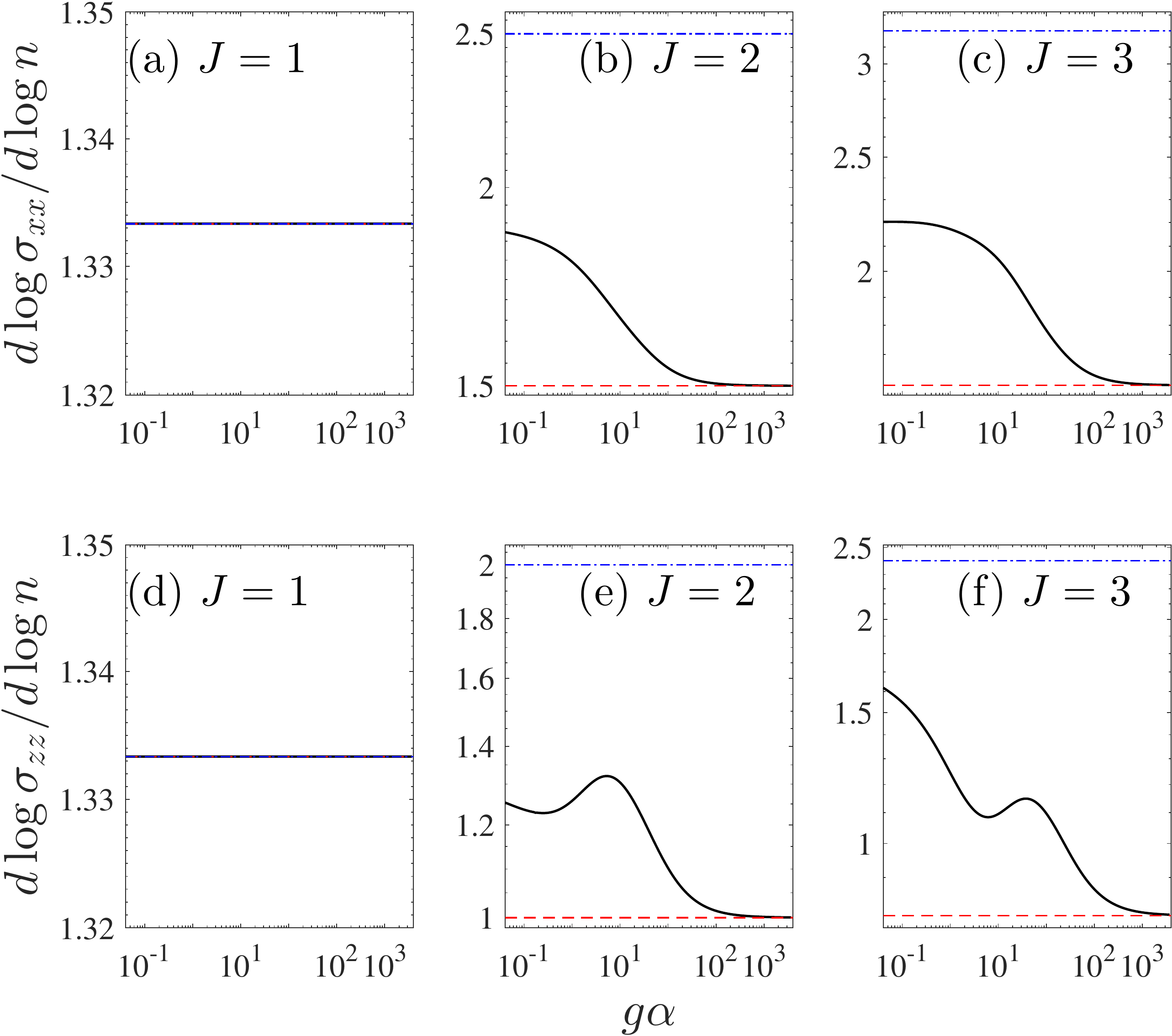}
\caption{
(a)-(c) $d\log \sigma_{xx}/d\log n$ and (d)-(f) $d\log \sigma_{zz}/d\log n$ as a function of the screening strength $g\alpha$ for charged impurities. Red dashed and blue dashed-dotted lines represent the density exponents obtained from $\zeta={1\over J}$ (or in the strong screening limit) and $\zeta=1$
in Eq.~(\ref{eq:density_dependence_charged_weak_screening}), respectively. Here, $n=n_0$ is used for the calculation.
}
\label{fig:density_power_evolution}
\end{figure}

Figure \ref{fig:density_power_evolution} illustrates the evolution of the power-law density dependence of the dc conductivity as a function of the screening strength characterized by $g\alpha$. Note that $\zeta={1\over J}$ in Eq.~(\ref{eq:density_dependence_charged_weak_screening}) gives the same density exponent as in the strong screening limit in Eq.~(\ref{eq:density_dependence_charged_strong_screening}).
Thus, as $\alpha$ increases, the density exponent evolves from that obtained in Eq.~(\ref{eq:density_dependence_charged_weak_screening}) with decreasing $\zeta$ within the range ${1\over J} \le \zeta \le 1$. Here, nonmonotonic behavior in the density exponent  originates from the angle-dependent power law in the relaxation time, which manifests in the weak screening limit. (See Sec.~\ref{sec:conductivity_3D_anisotropic} in SM \cite{SM} for further discussion.)

Similarly, for short-ranged impurities, $V_{\rm F}$ is a constant independent of density; in this case we find
\begin{subequations}\label{eq:density_dependence_short}
\begin{eqnarray}
\sigma_{xx}&\sim& \varepsilon_{\rm F}^{2\left(1-{1\over J}\right)}\sim n^{2(J-1)\over J+2}, \\
\sigma_{zz}&\sim& \varepsilon_{\rm F}^0 \sim n^0.
\end{eqnarray}
\end{subequations}

\begin{figure}[htb]
\includegraphics[width=1\linewidth]{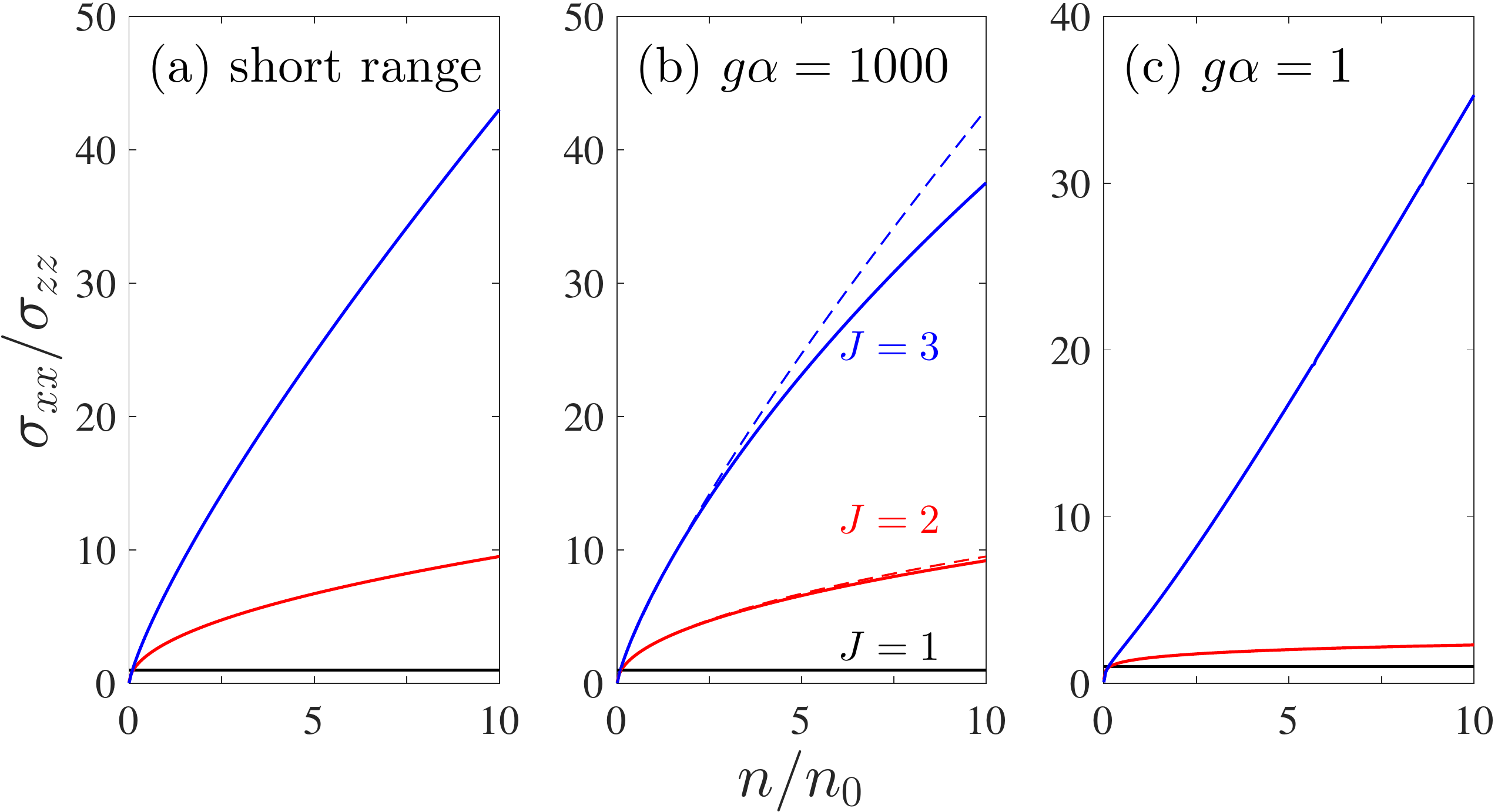}
\caption{
$\sigma_{xx}/\sigma_{zz}$ as a function of density for m-WSMs with $J=1,2,3$ for (a) short-range impurities, (b) charged impurities with $g\alpha=1000$, and (c) charged impurities with $g\alpha=1$. Dashed lines in (b) represent analytic forms in the strong screening limit given by Eq.~(\ref{eq:sigma_strong_screening_analytic}) in SM \cite{SM}.
}
\label{fig:anisotropy_factor}
\end{figure}

The anisotropy in conductivity can be characterized by $\sigma_{xx}/\sigma_{zz}$. Figure \ref{fig:anisotropy_factor} shows $\sigma_{xx}/\sigma_{zz}$ as a function of density for m-WSMs. Thus, as the carrier density increases, the anisotropy in conductivity increases. Interestingly, $\sigma_{xx}/\sigma_{zz}$ for both short-range impurities and charged impurities in the strong screening limit is given by
\begin{equation}\label{eq:anisotropy_factor}
\sigma_{xx}/\sigma_{zz}\sim \varepsilon_{\rm F}^{2\left(1-{1\over J}\right)}\sim n^{2(J-1)\over J+2}.
\end{equation}
Note that for arbitrary screening, $\zeta$s for $\sigma_{xx}$ and $\sigma_{zz}$ in Eq.~(\ref{eq:density_dependence_charged_weak_screening}) are actually different, thus not cancelled in $\sigma_{xx}/\sigma_{zz}$ and the power-law deviates from that in Eq.~(\ref{eq:anisotropy_factor}). (See Sec.~\ref{sec:conductivity_3D_anisotropic} in SM \cite{SM} for the analytic/asymptotic expressions of the density dependence of $\sigma_{xx}/\sigma_{zz}$.)

We consider both the short-range and charged impurities by adding their scattering rates according to Matthiessen's rule assuming that each scattering mechanism is independent.  At low densities (but high enough to validate the Boltzmann theory) the charged impurity scattering always dominates the short-range scattering, while at high densities the short-range scattering dominates, irrespective of the chirality $J$ and screening strength.

{\em Temperature dependence of dc conductivity}. 
In 3D materials, it is not easy to change the density of charge carriers by gating, because of screening in the bulk. However, the temperature dependence of dc conductivity can be used to understand the carrier dynamics of the system.
The effect of finite temperature arises from the energy averaging over the Fermi distribution function in Eq.~(\ref{eq:conductivity_tensor}), and the temperature dependence of the screening of the impurity potential for charged impurities \cite{Ando1982,DasSarma2011}.

From the invariance of carrier density with respect to temperature, we obtain the variation of the chemical potential $\mu(T)$ as a function of temperature $T$. Then the Thomas-Fermi wavevector $q_{\rm TF}(T)$ in 3D at finite $T$ can be expressed as $q_{\rm TF}(T)=\sqrt{{4\pi e^2\over \kappa} {\partial n \over \partial \mu}}$. 
In the low- and high-temperature limits, the chemical potential is given by
\begin{eqnarray}\label{eq:asymptotic_mu}
{\frac{\mu}{\varepsilon_{\rm F}}}\!\! &=&\!\!
\begin{cases}
1- \frac{\pi^2}{3J} \left(\frac{T}{T_{\rm F}} \right)^2 & (T\ll T_{\rm F}), \\ {1\over 2\eta\left({2\over J}\right)\Gamma\left(2+{2\over J}\right)}\left(\frac{T}{T_{\rm F}}\right)^{-{2\over J}} & (T\gg T_{\rm F}),
\end{cases}
\end{eqnarray}
whereas the Thomas-Fermi wave vector is given by
\begin{eqnarray}\label{eq:asymptotic_qtf}
\!\!\!{\frac{q_{\rm{TF}}(T) }{q_{\rm{TF}}(0) }}\!\! &= &\!\!
\begin{cases}
1- \frac{\pi^2}{6J} \left(\frac{T}{T_{\rm F}} \right)^2  & \!\!\!(T\ll T_{\rm F}), \\
\sqrt{2\eta\left({2\over J}\right)\Gamma\left(1+{2\over J}\right)}\left(\frac{T}{T_{\rm F}}\right)^{1\over J} & \!\!\!(T\gg T_{\rm F}),
\end{cases}
\end{eqnarray}
where $T_{\rm F}=\varepsilon_{\rm F}/k_{\rm B}$ is the Fermi temperature, and $\Gamma$ and $\eta$ are the gamma function and the Dirichlet eta function \cite{Arfken2012}, respectively.
(See Sec.~\ref{sec:chemical_potential_screening_wavevector} in SM \cite{SM} for the temperature dependence of the chemical potential and Thomas-Fermi wave vector.)
In a single-band system, $q_{\rm TF}(T)$ always decreases with $T^{-1}$ at high temperatures, whereas in m-WSMs, $q_{\rm TF}(T)$ increases with $T^{1\over J}$ because of the thermal excitation of carriers that participate in the screening.

\begin{figure}[htb]
\includegraphics[width=1\linewidth]{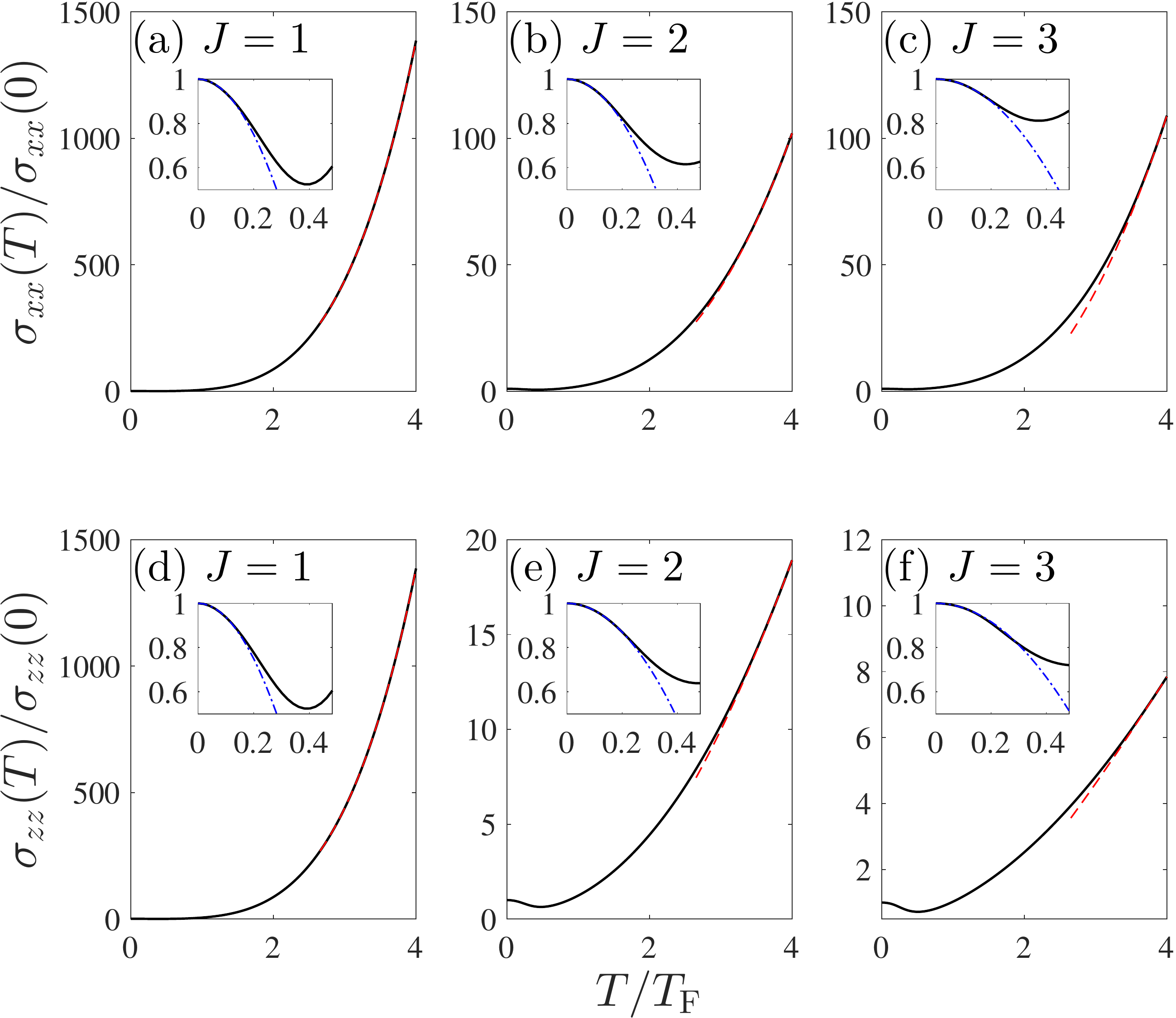}
\caption{
Temperature dependence of dc conductivity (a)-(c) $\sigma_{xx}$ and (d)-(f) $\sigma_{zz}$ for charged impurities with $g\alpha=1000$. The insets in each panel show the low temperature behavior. Red dashed and blue dashed-dotted lines represent fitting by Eq.~(\ref{eq:temperature_dependence_charged}) with $\zeta={1\over J}$ in the high- and low- temperature limits, respectively. 
}
\label{fig:temperature_dependence_charged}
\end{figure}

Figure \ref{fig:temperature_dependence_charged} shows the temperature dependence of dc conductivity for charged impurities. 
We find
\begin{subequations}\label{eq:temperature_dependence_charged}
\begin{eqnarray}
{\sigma_{xx}(T)\over \sigma_{xx}(0)}\!\! &=&\!\!
\begin{cases}
1+C_{xx}\left({T\over T_{\rm F}}\right)^2 & (T\ll T_{\rm F}), \\
D_{xx}\left({T\over T_{\rm F}}\right)^{2+4\zeta-{2\over J}} & (T\gg T_{\rm F}),
\end{cases} \\
{\sigma_{zz}(T)\over \sigma_{zz}(0)}\!\! &=&\!\!
\begin{cases}
1+C_{zz}\left({T\over T_{\rm F}}\right)^2 & (T\ll T_{\rm F}), \\
D_{zz}\left({T\over T_{\rm F}}\right)^{4\zeta} & (T\gg T_{\rm F}).
\end{cases}\end{eqnarray}
\end{subequations}
As discussed, $\zeta$ varies within ${1\over J}\le \zeta \le 1$ and approaches ${1\over J}$ in the strong screening limit ($g\alpha\gg 1$). Here, the high-temperature coefficients $D_{ii}>0$, whereas the low-temperature coefficients $C_{ii}$ change sign from negative to positive as $\alpha$ increases.
For short-range impurities, we find
\begin{subequations}\label{eq:temperature_dependence_short}
\begin{eqnarray}
{\sigma_{xx}(T)\over \sigma_{xx}(0)}\!\! &=&\!\!
\begin{cases}
1+C_{xx}^{\rm short}\left({T\over T_{\rm F}}\right)^2 & (T\ll T_{\rm F}), \\
D_{xx}^{\rm short}\left({T\over T_{\rm F}}\right)^{2(J-1)\over J} & (T\gg T_{\rm F}),
\end{cases} \\
{\sigma_{zz}(T)\over \sigma_{zz}(0)}\!\! &=&\!\!
\begin{cases}
1-e^{-T_{\rm F}/T} &\! (T\ll T_{\rm F}), \\
{1\over 2}+D_{zz}^{\rm short}\left({T\over T_{\rm F}}\right)^{-{2+J\over J}} &\! (T\gg T_{\rm F}).
\end{cases}
\end{eqnarray}
\end{subequations}
Here, $C_{xx}^{\rm short}<0$ and $D_{ii}^{\rm short}>0$. 
Note that for $J=1$, Eq.~(\ref{eq:temperature_dependence_short}a) becomes constant, and reduces to Eq.~(\ref{eq:temperature_dependence_short}b) if next order corrections are included. 
(See Sec.~\ref{sec:temperature_dependence} in SM \cite{SM} for the analytic/asymptotic expressions of the temperature coefficients, and the evolution of $C_{ii}$ as a function of $g\alpha$.)

To understand the temperature dependence, we can consider a situation where the thermally induced charge carriers participate in transport. Then the temperature dependence in the high-temperature limit can be obtained simply by replacing the $\varepsilon_{\rm F}$ dependence with $T$ in Eqs.~(\ref{eq:density_dependence_charged_strong_screening})-(\ref{eq:density_dependence_short}), which describe the density dependence of dc conductivity. Similarly as in Fig.~\ref{fig:anisotropy_factor}, $\sigma_{xx}(T)/\sigma_{zz}(T)$ also increases with $T$ at high temperatures.

For the charged impurities at high temperatures, and neglecting the effect of phonons, the conductivity increases with temperature, and mimics an insulating behavior.
By contrast, for short-range impurities at high temperatures, $\sigma_{zz}(T)$ decreases with temperature and approaches $0.5 \sigma_{zz}(0)$, thus showing a metallic behavior. Interestingly, $\sigma_{xx}(T)$  shows contrasting behavior for $J>1$ and $J=1$, increasing (decreasing) with temperature for $J>1$ ($J=1$) showing insulating (metallic) behavior at high temperatures.

{\em Discussion}. 
We find that the dc conductivities in the Boltzmann limit show characteristic density and temperature dependences that depend strongly on the chirality of the system, revealing a signature of m-WSMs in transport measurements, which can be compared with experiments. In real materials with time reversal symmetry, multiple Weyl points with compensating chiralities will be present. The contributions from the individual nodes calculated by our  method are additive when the Weyl points are well separated and internode scattering is weak.
Our analysis is based on the semiclassical Boltzmann transport theory with the Thomas-Fermi approximation for screening and corrected for the anisotropy of the Fermi surface in m-WSMs. The Boltzmann transport theory is known to be valid in the high density limit. At low densities, inhomogeneous impurities induce a spatially varying local chemical potential, typically giving a minimum conductivity when the chemical potential is  at the Weyl node \cite{Ramakrishnan2015} and the problem is treated within the effective medium theory. Note that the Thomas-Fermi approximation used in this work is the long-wavelength limit of the random phase approximation (RPA), and neglects interband contributions to the polarization function \cite{Ramakrishnan2015}, thus deviating from the RPA result at low densities. Both simplifications become important in the low-density limit, which will be considered in our future work.  \\


The authors thank Shaffique Adam for helpful discussions and comments.
This research was supported by the Basic Science Research Program through the National Research Foundation of Korea (NRF) funded by the Ministry of Education under Grant No. 2015R1D1A1A01058071.
E.J.M.'s work on this project was supported by the U.S. Department of Energy, Office of Basic Energy Sciences under Award No. DE-FG02-ER45118.
H.M. acknowledges travel support provided by the University Research Foundation at the University of Pennsylvania while this work was carried out.



\clearpage 
\widetext
\setcounter{section}{0}
\setcounter{equation}{0}
\setcounter{figure}{0} 
\setcounter{table}{0} 
\renewcommand\thefigure{S\arabic{figure}} 
\setcounter{page}{1}

\large
\begin{center}
{\bf Supplemental Material:\\
Semiclassical Boltzmann transport theory for multi-Weyl semimetals}
\end{center}
\normalsize

%

\section{Eigenstates and density of states for multi-Weyl semimetals}
\label{sec:eigenstate_dos}
Let us consider the eigenstates and density of states (DOS) for the low-energy effective Hamiltonian of m-WSMs described by Eq.~(\ref{eq:wsm_J}) in the main text:
\begin{eqnarray}
H_J
&=&\varepsilon_0\left(
\begin{array}{cc}
c_z \tilde{k}_z & \tilde{k}_{-}^J\\
\tilde{k}_{+}^J &-c_z \tilde{k}_z\\
\end{array}
\right),
\end{eqnarray}
where $\tilde{\bm{k}}=\bm{k}/k_0$ and $c_z=\hbar v_z k_0/\varepsilon_0$. To avoid difficulties associated with anisotropic dispersions, we consider the following coordinate transformation \cite{Ahn2016_SM}
\begin{equation}\label{eq:transformation}
\begin{split}
k_x&\rightarrow k_0\left(r\sin\theta\right)^{1\over J}\cos\phi,\\
k_y&\rightarrow k_0\left(r\sin\theta\right)^{1\over J}\sin\phi,\\
k_z&\rightarrow {k_0 \over c_z} r\cos\theta,
\end{split}
\end{equation}
which transforms the Hamiltonian into the following form:
\begin{equation}\label{anisotropic_model}
H=\varepsilon_0 r \left(
\begin{array}{cc}
\cos\theta &\sin\theta e^{-iJ\phi}\\
\sin\theta e^{iJ\phi} &-\cos\theta\\
\end{array}
\right).
\end{equation}
In the transformed coordinates, the energy dispersion is given by $\varepsilon_{\pm}(r)=\pm \varepsilon_0 r$ and the corresponding eigenstate is given by
\begin{subequations}
\begin{eqnarray}
|+\rangle&=&
\left(
\begin{array}{c}
\cos{\theta\over 2} \\
\sin{\theta\over 2} e^{i J\phi} \\
\end{array}
\right), \\
|-\rangle&=&
\left(
\begin{array}{c}
-\sin{\theta\over 2} \\
\cos{\theta\over 2} e^{i J\phi}
\end{array}
\right).
\end{eqnarray}
\end{subequations}

The Jacobian $\mathcal{J}$ corresponding to this transformation is given by
\begin{equation}\label{eq:jacobian}
\mathcal{J}=
\left|
\begin{array}{ccc}
 \frac{\partial k_x}{\partial r} & \frac{\partial k_x}{\partial \theta } & \frac{\partial k_x}{\partial \phi} \\
 \frac{\partial k_y}{\partial r} & \frac{\partial k_y}{\partial \theta } & \frac{\partial k_y}{\partial \phi} \\
 \frac{\partial k_z}{\partial r} & \frac{\partial k_z}{\partial \theta } & \frac{\partial k_z}{\partial \phi} \\
\end{array}
\right|={k_0^3 \over c_z J} r^{2\over J}\sin^{{2\over J}-1}\theta \equiv \mathcal{J}(r,\theta).
\end{equation}

Note that for the $+$ band, the band velocity $v_{\bm k}^{(i)}={1\over \hbar} {\varepsilon_{+,\bm{k}} \over \partial  k_i}$ can be expressed as
\begin{subequations}\label{eq:band_velocity}
\begin{eqnarray}
v_{\bm k}^{(x)}&=&J v_0 r^{1-{1\over J}} \sin^{2-{1\over J}}\theta \cos\phi, \\
v_{\bm k}^{(y)}&=&J v_0 r^{1-{1\over J}} \sin^{2-{1\over J}}\theta \sin\phi, \\
v_{\bm k}^{(z)}&=&c_z v_0 \cos\theta,
\end{eqnarray}
\end{subequations}
where $v_0={\varepsilon_0\over \hbar k_0}$.

The DOS at energy $\varepsilon>0$ can be obtained as
\begin{eqnarray}\label{eq:dos_m-WSM}
D(\varepsilon)&=&g\int {d^3k\over (2\pi)^3} \delta(\varepsilon-\varepsilon_{+,\bm{k}}) \nonumber \\
&=&g \int_0^{\infty}dr \int_0^{\pi}d\theta \int_0^{2\pi}d\phi {\mathcal{J}(r,\theta)\over (2\pi)^3}\delta(\varepsilon-\varepsilon_0 r) \nonumber \\
&=&{g B\left({1\over 2},{1\over J}\right)\over 4\pi^2 c_z J}{k_0^3\over \varepsilon_0}\left(\varepsilon\over \varepsilon_0\right)^{2\over J},
\end{eqnarray}
where $g$ is the number of degenerate Weyl nodes.
Here, we used the relation $\int_0^{\pi/2} d\theta \cos^m\theta \sin^n\theta={1\over 2} B({m+1\over 2},{n+1\over 2})$, where $B(m,n)={\Gamma(m)\Gamma(n) \over \Gamma(m+n)}$ is the beta function and $\Gamma(x)=\int_0^{\infty} dt\, t^{x-1}e^{-t}$ is the gamma function \cite{Arfken_SM}.
Note that the Thomas-Fermi wavevector is determined by the DOS at the Fermi energy $\varepsilon_{\rm F}$ given by
\begin{equation}\label{eq:q_TF_mWSM}
q_{\rm TF}=\sqrt{{4\pi e^2 \over \kappa} D(\varepsilon_{\rm F})}=k_0 \sqrt{g \alpha B\left({1\over 2},{1\over J}\right)\over \pi c_z J}\left(\varepsilon_{\rm F}\over \varepsilon_0\right)^{1\over J},
\end{equation}
where $\alpha={e^2\over \kappa \hbar v_0}$ is the effective fine structure constant.

The carrier density is then given by
\begin{equation}
n=\int_0^{\varepsilon_{\rm F}}d\varepsilon D(\varepsilon)=n_0 {g B\left({1\over 2},{1\over J}\right)\over 4\pi^2 c_z (J+2)}\left(\varepsilon_{\rm F}\over \varepsilon_0\right)^{{2\over J}+1},
\end{equation}
where $n_0=k_0^3$.
Note that $\varepsilon_{\rm F}\sim n^{J\over  J+2}$ and $D(\varepsilon_{\rm F})\sim n^{2\over J+2}$.

\section{Density dependence of dc conductivity in multi-Weyl semimetals at zero temperature}
\label{sec:conductivity_3D_anisotropic}

In this section, we derive the dc conductivity at zero temperature for 3D anisotropic systems with an anisotropic energy dispersion which has an axial symmetry around the $k_z$-axis (i.e. independent of $\phi$), as in the m-WSMs described by Eq.~(\ref{eq:wsm_J}) in the main text. To take into account the anisotropy of the energy dispersion, we express the anisotropic Boltzmann equation in Eq.~(\ref{eq:relaxation_time_anisotropic}) in the main text using the transformed coordinates in Eq.~(\ref{eq:jacobian}) assuming an axial symmetry around the $k_z$-axis:
\begin{eqnarray}\label{eq:relaxation_time_anisotropic_theta}
1&=&\int_0^{\infty}dr'\int_0^{\pi}d\theta'\int_0^{2\pi}d\phi'{\mathcal{J}(r',\theta')\over (2\pi)^3} W_{\bm{k}\bm{k}'} \left(\tau_{\bm{k}}^{(i)}-{v_{\bm{k}'}^{(i)}\over v_{\bm{k}}^{(i)} }\tau_{\bm{k}'}^{(i)}\right) \nonumber \\
&=&\int_0^{\infty}dr'\int_0^{\pi}d\theta'\int_0^{2\pi}d\phi'
{k_0^3 {r'}^{2\over J}\sin^{{2\over J}-1}\theta'\over (2\pi)^3 c_z J}
\left[{2\pi\over\hbar} n_{\rm imp} |V_{\bm{k}\bm{k}'}|^2 F_{\bm{k}\bm{k}'} \delta(\varepsilon_0 r-\varepsilon_0 r')\right]\left(\tau_{\bm{k}}^{(i)}-d_{\bm{k}\bm{k}'}^{(i)}\tau_{\bm{k}'}^{(i)}\right) \nonumber \\
&=&{2\pi\over\hbar} n_{\rm imp} {k_0^3 r^{2\over J} \over (2\pi)^2 c_z J \varepsilon_0}\int_{-1}^{1}d\cos\theta' (1-\cos^2\theta')^{{1\over J}-1} \int_0^{2\pi}{d\phi'\over 2\pi} |V_{\bm{k}\bm{k}'}|^2 F_{\bm{k}\bm{k}'} \left(\tau_{\bm{k}}^{(i)}-d_{\bm{k}\bm{k}'}^{(i)}\tau_{\bm{k}'}^{(i)}\right),
\end{eqnarray}
where $d_{\bm{k}\bm{k}'}^{(i)}=v_{\bm{k}'}^{(i)}/v_{\bm{k}}^{(i)}$ and $F_{\bm{k}\bm{k}'} =  \frac{1}{2} \left[1+\cos\theta \cos\theta' +\sin \theta \sin\theta' \cos J(\phi - \phi')\right]$ is the square of the wavefunction overlap between $\bm{k}$ and $\bm{k}'$ states in the same band.
Let us define $\rho_0={k_0^3 \over (2\pi)^2  c_z \varepsilon_0}$, $V_0={\varepsilon_0\over k_0^3}$, and ${1\over \tau_0(r)}={2\pi\over \hbar} n_{\rm imp} V_0^2 \rho_0$. Then with $\mu=\cos\theta$, we have
\begin{equation}\label{eq:relaxation_time_anisotropic_mu}
1={r^{2\over J}\over J} \int_{-1}^{1}d\mu'(1-{\mu'}^2)^{{1\over J}-1} \int_0^{2\pi}{d\phi'\over 2\pi} |\tilde{V}_{\bm{k}\bm{k}'}|^2 F_{\bm{k}\bm{k}'} \left(\tilde{\tau}_{\bm{k}}^{(i)}-d_{\bm{k}\bm{k}'}^{(i)}\tilde{\tau}_{\bm{k}'}^{(i)}\right),
\end{equation}
where $\tilde{V}_{\bm{k}\bm{k}'}=V_{\bm{k}\bm{k}'}/V_0$ and $\tilde{\tau}_{\bm{k}}^{(i)}=\tau_{\bm{k}}^{(i)}/\tau_0$.


Assuming $\tilde{\tau}_{\bm{k}}^{(i)}=\tilde{\tau}^{(i)}(\mu)$ from the axial symmetry,
\begin{equation}\label{eq:Boltzmann_mWSM}
1=\tilde{w}^{(i)}(\mu)\tilde{\tau}^{(i)}(\mu)-\int_{-1}^{1}d\mu' \tilde{w}^{(i)}(\mu,\mu')\tilde{\tau}^{(i)}(\mu'),
\end{equation}
where
\begin{subequations}
\label{eq:scattering_rate_matrix}
\begin{eqnarray}
\tilde{w}^{(i)}(\mu)&=&{r^{2\over J}\over J}\int_{-1}^{1}d\mu'(1-{\mu'}^2)^{{1\over J}-1} \int_0^{2\pi}{d\phi'\over 2\pi} |\tilde{V}_{\bm{k}\bm{k}'}|^2 F_{\bm{k}\bm{k}'}, \\
\tilde{w}^{(i)}(\mu,\mu')&=&{r^{2\over J}\over J}(1-{\mu'}^2)^{{1\over J}-1} \int_0^{2\pi}{d\phi'\over 2\pi} |\tilde{V}_{\bm{k}\bm{k}'}|^2 F_{\bm{k}\bm{k}'} d_{\bm{k}\bm{k}'}^{(i)}.
\end{eqnarray}
\end{subequations}

Now let us discretize $\theta$ or equivalently $\mu=\cos\theta$ to $\mu_n$ ($n=1,2,\cdots,N$) with an interval $\Delta\mu=2/N$. Then for $\tilde{\tau}_n^{(i)}=\tilde{\tau}^{(i)}(\mu_n)$, we have
\begin{equation}
\label{eq:relaxation_time_matrix_form}
1=P_n^{(i)} \tilde{\tau}_n^{(i)} - \sum_{n'} P_{nn'}^{(i)}\tilde{\tau}_{n'}^{(i)},
\end{equation}
where $P_n^{(i)}=\tilde{w}^{(i)}(\mu_n)$ is an $N$-vector and $P_{nn'}^{(i)}=\tilde{w}^{(i)}(\mu_n,\mu_{n'})\Delta\mu$ is an $N\times N$ matrix which relate the $\theta$-dependent relaxation times. Note that Eq.~(\ref{eq:relaxation_time_matrix_form}) has a similar structure for the multiband scattering \cite{Siggia1970_SM} in which the relaxation time can be obtained by solving coupled equations, which relate the relaxation times for different energy bands involved in the scattering.

Then the dc conductivity at zero temperature is given by
\begin{eqnarray}
\label{eq:conductivity_zero_temperature}
\sigma_{ij}&=&g e^2\int {d^3 k\over (2\pi)^3} \delta(\varepsilon_{\bm k}-\varepsilon_{\rm F}) v_{\bm{k}}^{(i)} v_{\bm{k}}^{(j)}\tau_{\bm{k}}^{(j)} \nonumber \\
&=&g e^2 \int_0^{\infty}dr \int_0^{\pi}d\theta \int_0^{2\pi}d\phi {k_0^3 r^{2\over J}\sin^{{2\over J}-1}\theta\over (2\pi)^3 c_z J}
\delta(\varepsilon_0 r-\varepsilon_{\rm F}) v_{\bm{k}}^{(i)} v_{\bm{k}}^{(j)}\tau_{\bm{k}}^{(j)} \nonumber \\
&=&{\sigma_0 \over J} \int_0^{\infty}dr r^{2\over J} \int_{-1}^{1}d\mu \left(1-\mu^2\right)^{{1\over J}-1} \int_0^{2\pi}{d\phi\over 2\pi} \delta(r-r_{\rm F}) \tilde{v}_{\bm{k}}^{(i)} \tilde{v}_{\bm{k}}^{(j)} \tilde{\tau}_{\bm{k}}^{(j)},
\end{eqnarray}
where $\sigma_0=g e^2 \rho_0 v_0^2 \tau_0$, $r_{\rm F}=\varepsilon_{\rm F}/\varepsilon_0$,  and $\tilde{v}_{\bm{k}}^{(i)}=v_{\bm{k}}^{(i)}/v_0$. Thus, from Eq.~(\ref{eq:band_velocity}), we have
\begin{subequations}\label{eq:sigma_xx_sigma_zz}
\begin{eqnarray}
{\sigma_{xx} \over \sigma_0} &=& {J r_{\rm F}^2\over 2} \int_{-1}^{1}d\mu \left(1-\mu^2\right)\tilde{\tau}^{(x)}(\mu), \\
{\sigma_{zz} \over \sigma_0} &=& {c_z^2 r_{\rm F}^{2\over J}\over J} \int_{-1}^{1}d\mu \left(1-\mu^2\right)^{{1\over J}-1}\mu^2 \tilde{\tau}^{(z)}(\mu).
\end{eqnarray}
\end{subequations}
Note that $\tau_0$, $v_0$, $\rho_0$ and $\sigma_0$ are the density independent normalization constants in units of time, velocity, DOS, and conductivity, respectively. In addition, from the axial symmetry around the $k_z$-axis, $\sigma_{xx}=\sigma_{yy}$.

For the short-range impurities, $V_{\bm{k}\bm{k}'}$ is independent of density. Thus from Eq.~(\ref{eq:scattering_rate_matrix}), $\tilde{\omega}^{(i)}(\mu)\sim
\varepsilon_{\rm F}^{2\over J}$ and $\tilde{\tau}^{(i)}(\mu)\sim \varepsilon_{\rm F}^{-{2\over J}}$ at the Fermi energy $\varepsilon_{\rm F}$. Note that $\varepsilon_{\rm F}\sim n^{J\over J+2}$. Therefore we have
\begin{subequations}\label{eq:sigma_density_dependence_short_range}
\begin{eqnarray}
\sigma_{xx}&\sim& \varepsilon_{\rm F}^{2-{2\over J}} \sim n^{2(J-1)\over J+2}, \\
\sigma_{zz}&\sim& \varepsilon_{\rm F}^0 \sim n^0.
\end{eqnarray}
\end{subequations}

For charged impurities in the strong screening limit, $V_{\bm{k}\bm{k}'}\sim q_{\rm TF}^{-2}\sim D^{-1}(\varepsilon_{\rm F})\sim \varepsilon_{\rm F}^{-{2\over J}}$, thus $\tilde{\omega}^{(i)}(\mu)\sim\varepsilon_{\rm F}^{{2\over J}-{4\over J}}$ and $\tilde{\tau}^{(i)}(\mu)\sim \varepsilon_{\rm F}^{2\over J}$ at $\varepsilon_{\rm F}$. Therefore we have
\begin{subequations}\label{eq:sigma_density_dependence_strong_screening}
\begin{eqnarray}
\sigma_{xx}&\sim& \varepsilon_{\rm F}^{2+{2\over J}} \sim n^{2(J+1)\over J+2}, \\
\sigma_{zz}&\sim& \varepsilon_{\rm F}^{4\over J} \sim n^{4\over J+2}.
\end{eqnarray}
\end{subequations}

For charged impurities in the weak screening limit, from $V_{\bm{k}\bm{k}'}\sim |\bm{k}-\bm{k}'|^{-2}$ and Eq.~(\ref{eq:transformation}), we expect the potential average on the Fermi surface as $V_{\rm F}\sim \varepsilon_{\rm F}^{-2\zeta}$ with ${1\over J} \le \zeta \le 1$ (assuming no logarithmic correction), thus $\tilde{\omega}^{(i)}(\mu)\sim\varepsilon_{\rm F}^{{2\over J}-4\zeta}$ and $\tilde{\tau}^{(i)}(\mu)\sim \varepsilon_{\rm F}^{4\zeta-{2\over J}}$ at $\varepsilon_{\rm F}$. Therefore, we have
\begin{subequations}\label{eq:sigma_density_dependence_weak_screening}
\begin{eqnarray}
\sigma_{xx}&\sim& \varepsilon_{\rm F}^{2+4\zeta-{2\over J}} \sim n^{2(J-1)+4J\zeta \over J+2}, \\
\sigma_{zz}&\sim& \varepsilon_{\rm F}^{4\zeta}\sim n^{4J\zeta \over J+2}.
\end{eqnarray}
\end{subequations}
Here, $\zeta$s in $\sigma_{xx}$ and $\sigma_{zz}$ do not need to be the same, as explained later in this section. Note that $\zeta={1\over J}$ gives the same density exponent corresponding to the strong screening limit.  

For the short-range impurities, it turns out that the relaxation time is independent of polar angles $\theta$. Assuming $\tau^{(i)}(\mu)=\tau^{(i)}$ from the beginning, for short-range impurity potential $V_{\bm{k}\bm{k}'}=V_{\rm short}$, Eq.~(\ref{eq:relaxation_time_anisotropic_mu}) reduces to
\begin{equation}\label{eq:relaxation_time_anisotropic_mu_theta_independent}
{1\over \tilde{\tau}^{(i)}}={r^{2\over J}\over J} \tilde{V}_{\rm short}^2  \int_{-1}^{1}d\mu'(1-{\mu'}^2)^{{1\over J}-1} \int_0^{2\pi}{d\phi'\over 2\pi}  F_{\bm{k}\bm{k}'} \left(1-d_{\bm{k}\bm{k}'}^{(i)}\right),
\end{equation}
where $\tilde{V}_{\rm short}=V_{\rm short}/V_0$.
Then we find that the relaxation time $\tau^{(i)}(\varepsilon)$ at at energy $\varepsilon=r\varepsilon_0$ is
\begin{subequations}\label{eq:tau_short_analytic}
\begin{eqnarray}
{1\over \tilde{\tau}^{(x)}(\varepsilon)}&=& {r^{2\over J}\over 2J} \tilde{V}_{\rm short}^2 B\left({1\over 2},{1\over J}\right)-\delta_{J1}{r^2\over 3}\tilde{V}_{\rm short}^2, \\
{1\over \tilde{\tau}^{(z)}(\varepsilon)}&=& {r^{2\over J}\over 2J} \tilde{V}_{\rm short}^2 B\left({1\over 2},{1\over J}+1\right).
\end{eqnarray}
\end{subequations}
From Eq.~(\ref{eq:sigma_xx_sigma_zz}), finally we obtain
\begin{subequations}\label{eq:sigma_short_analytic}
\begin{eqnarray}
{\sigma_{xx}\over \sigma_0}&=& {2J r_{\rm F}^2 \over 3} \tilde{\tau}^{(x)}(\varepsilon_{\rm F}), \\
{\sigma_{zz}\over \sigma_0}&=& {c_z^2 r_{\rm F}^{2\over J} \over J} B\left({3\over 2},{1\over J}\right)  \tilde{\tau}^{(z)}(\varepsilon_{\rm F})={J c_z^2 \over \tilde{V}_{\rm short}^2}.
\end{eqnarray}
\end{subequations}
Note that $\sigma_{xx}/\sigma_0=\tilde{V}_{\rm short}^{-2}$, ${16 \over 3\pi } r_{\rm F} \tilde{V}_{\rm short}^{-2}$, ${12 \over B\left({1\over 2},{1\over 3}\right)} r_{\rm F}^{4\over 3} \tilde{V}_{\rm short}^{-2}$ for $J=1,2,3$, respectively, and the obtained analytic expressions are consistent with the density dependence in Eq.~(\ref{eq:sigma_density_dependence_short_range}). From Eq.~(\ref{eq:sigma_short_analytic}), we find 
$\sigma_{xx}/\sigma_{zz}={1\over c_z^2}$, ${8 r_{\rm F} \over 3\pi c_z^2}$, ${4 r_{\rm F}^{4\over 3} \over B\left({1\over 2},{1\over 3}\right) c_z^2}$ for $J=1,2,3$, respectively, and the anisotropy between $\sigma_{xx}$ and $\sigma_{zz}$ increases as the Fermi energy or the carrier density increases.

For charged impurities in the strong screening limit, the impurity potential becomes $V_{\bm{k}\bm{k}'}\approx V^{\rm strong}_{\rm screen} \equiv {4\pi e^2 \over {\kappa q_{\rm TF}^2}}$, having the same feature of the short-range impurity potential. Thus, the relaxation time is also independent of polar angles and similar analytic expressions can be obtained by replacing $\tilde{V}_{\rm short}$ by $\tilde{V}^{\rm strong}_{\rm screen}$ in Eqs.~(\ref{eq:tau_short_analytic}) and (\ref{eq:sigma_short_analytic}), where $\tilde{V}^{\rm strong}_{\rm screen}=V^{\rm strong}_{\rm screen}/V_0=4\pi\alpha k_0^2/q_{\rm TF}^2$. Then the relaxation time is given by
\begin{subequations}\label{eq:tau_strong_screening}
\begin{eqnarray}
{1\over \tilde{\tau}^{(x)}(\varepsilon)}&=& {8\pi^4 c_z^2 J \over g^2 B\left({1\over 2},{1\over J}\right)} r^{2\over J} r_{\rm F}^{-{4\over J}}-\delta_{J1}{4\pi^4 c_z^2 \over 3 g^2} r^2 r_{\rm F}^{-4}, \\
{1\over \tilde{\tau}^{(z)}(\varepsilon)}&=& {16\pi^4 c_z^2 J \over (J+2) g^2 B\left({1\over 2},{1\over J}\right)} r^{2\over J} r_{\rm F}^{-{4\over J}},
\end{eqnarray}
\end{subequations}
thus, in the strong screening limit, we obtain
\begin{subequations}\label{eq:sigma_strong_screening_analytic}
\begin{eqnarray}
{\sigma_{xx}\over \sigma_0}&=& {2J r_{\rm F}^2 \over 3} \tilde{\tau}^{(x)}(\varepsilon_{\rm F}), \\
{\sigma_{zz}\over \sigma_0}&=& {c_z^2 r_{\rm F}^{2\over J} \over J} B\left({3\over 2},{1\over J}\right)  \tilde{\tau}^{(z)}(\varepsilon_{\rm F})={g^2 B^2\left({1\over 2},{1\over J}\right) \over 16\pi^4 J} r_{\rm F}^{4\over J}.
\end{eqnarray}
\end{subequations}
Note that $\sigma_{xx}/\sigma_0={g^2\over 4\pi^4 c_z^2} r_{\rm F}^4$, ${g^2 \over 12\pi^3 c_z^2} r_{\rm F}^3$, ${g^2 B({1\over 2},{1\over 3})\over 12\pi^4 c_z^2} r_{\rm F}^{8\over 3}$ for $J=1,2,3$, respectively, and the obtained analytic expressions are consistent with the density dependence in Eq.~(\ref{eq:sigma_density_dependence_strong_screening}). Also note that $\sigma_{xx}/\sigma_{zz}$ has the same form with that obtained for short-range impurities.

\begin{figure}[htb]
\includegraphics[width=0.6\linewidth]{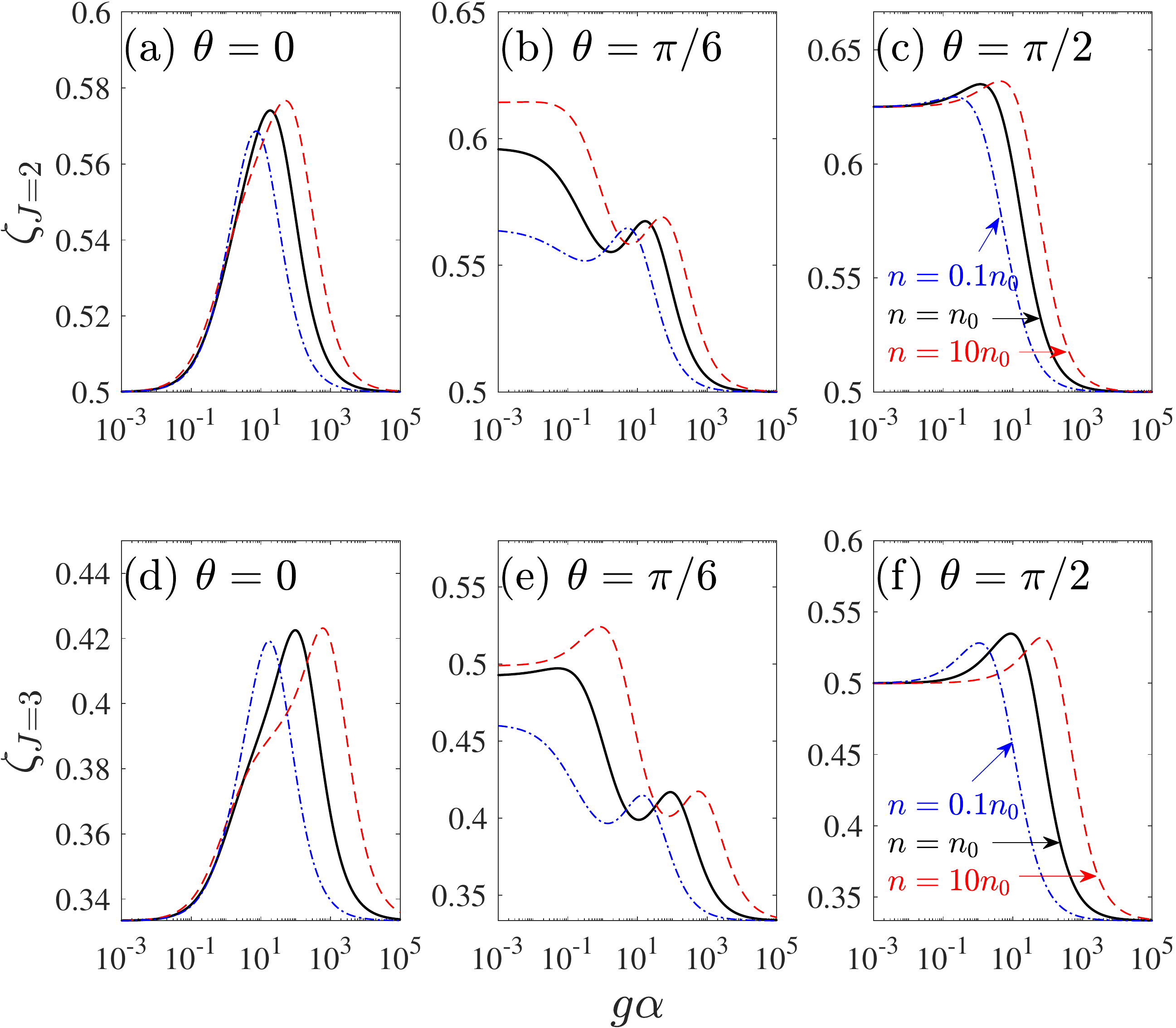}
\caption{
Angle dependent exponent $\zeta(\theta)$ for (a)-(c) $J=2$ and (d)-(f) $J=3$ as a function of the screening strength $g\alpha$ at $\theta=0, \pi/6, \pi/2$. Blue dashed-dotted, black solid, and red dashed lines represent $n=0.1 n_0, n_0, 10 n_0$, respectively.
}
\label{fig:angle_dependent_exponent}
\end{figure}

For charged impurities at arbitrary screening, the relaxation time in general depends on polar angles for $J>1$. In addition, as seen in Fig.~\ref{fig:density_power_evolution} in the main text, the density exponent shows non-monotonic behavior as a function of $g\alpha$. From Eq.~(\ref{eq:relaxation_time_anisotropic_theta}), for a given wavevector $\bm{k}=(k_0\left(r_{\rm F}\sin\theta\right)^{1\over J},0,{k_0 \over c_z} r_{\rm F}\cos\theta)$ at the Fermi energy, the average of the squared Coulomb potential on the Fermi surface is given by
\begin{equation}
\left<V^2(\theta)\right>_{\rm F}={1\over 2}\int_{-1}^{1}d\cos\theta' (1-\cos^2\theta')^{{1\over J}-1} \int_0^{2\pi}{d\phi'\over 2\pi} |V_{\bm{k}\bm{k}'}|^2.
\end{equation}
Then assuming $\left<V^2(\theta)\right>_{\rm F}\sim r_{\rm F}^{-4\zeta(\theta)}$, we can obtain the angle dependent exponent $\zeta(\theta)$ with ${1\over J} \le \zeta(\theta)\le 1$. Figure \ref{fig:angle_dependent_exponent} shows $\zeta(\theta)$ for several values of $\theta=0, \pi/6, \pi/2$. This angle-dependent power-law gives rise to a significant non-monotonic behavior of $\tau_z$ and $\sigma_{zz}$ in $g\alpha$, which originates from the competition between two inverse length scales, $q_{\rm TF}\sim r_{\rm F}^{1\over J}$ and $k_{\rm F}^{(z)}\sim r_{\rm F}$. Note that the in-plane component of the wavevector $k_{\rm F}^{(\parallel)}\sim r_{\rm F}^{1\over J}$ at the Fermi energy has the same Fermi energy dependence with $q_{\rm TF}$, showing a monotonic-like behavior of $\tau_x$ and $\sigma_{xx}$ in $g\alpha$. As $g\alpha$ increases, $\zeta(\theta)$ eventually  approaches $1/J$ irrespective of $\theta$, obtained in the strong screening limit.

\section{Temperature dependence of chemical potential and Thomas-Fermi wavevector in multi-Weyl semimetals}
\label{sec:chemical_potential_screening_wavevector}

In this section, we derive the temperature dependent chemical potential and Thomas-Fermi wavevector in a general gapless electron-hole system, and apply the results to m-WSMs. Suppose that a gapless electron-hole system has a DOS given by $D(\varepsilon) = C_\alpha |\varepsilon|^{\alpha-1}\varTheta(\varepsilon)$, where $C_\alpha$ is a constant and $\varTheta(\varepsilon)$ is a step function. For a $d$-dimensional electron gas with an isotropic energy dispersion $\varepsilon\sim k^J$, $\alpha=d/J$, whereas for m-WSMs, $D(\varepsilon) \propto \varepsilon^{2 \over J}$ from Eq.~(\ref{eq:dos_m-WSM}), thus $\alpha ={2\over J}+1$.

When the temperature is finite, the chemical potential $\mu$ deviates from the Fermi energy $\varepsilon_{\rm F}$ due to the broadening of the Fermi distribution function $f^{(0)}(\varepsilon,\mu)=\left[e^{\beta (\varepsilon-\mu)}+ 1\right]^{-1}$ where $\beta={1\over k_{\rm B}T}$. Since the charge carrier density $n$ does not vary under the temperature change, we have
\begin{equation}
n=\int_{-\infty}^{\infty}d\varepsilon D(\varepsilon)f^{(0)}(\varepsilon ,\mu)
=\int_{0}^{\infty}d\varepsilon D(\varepsilon)\left[f^{(0)}(\varepsilon ,\mu)+f^{(0)}(-\varepsilon ,\mu)\right] \equiv \int_{-\infty}^{\varepsilon_{\rm F}}d\varepsilon D(\varepsilon).
\end{equation}
Then the carrier density measured from the charge neutral point, $\Delta n\equiv \left.n\right|_{\mu}-\left.n\right|_{\mu=0}$, is given by
\begin{equation}
\Delta n= \int_{0}^{\infty}d\varepsilon D(\varepsilon)\left[f^{(0)}(\varepsilon ,\mu)-f^{(0)}(\varepsilon ,-\mu)\right] \equiv \int_{0}^{\varepsilon_{\rm F}}d\varepsilon D(\varepsilon).
\end{equation}
Here, we used $f(-\varepsilon,\mu)=1-f(\varepsilon,-\mu)$.

Before proceeding further, let us consider the following integral:
\begin{eqnarray}
\label{eq:fermionic_integral}
\int_0^{\infty}dx {x^{\alpha-1}\over z^{-1} e^x+1}
&=&\int_0^{\infty}dx {x^{\alpha-1} z e^{-x} \over 1+ze^{-x}} =-\int_0^{\infty}dx {x^{\alpha-1} \sum_{n=1}^{\infty} (-z)^n e^{-nx}} \nonumber \\
&\mathop{=}\limits^{t=nx}&\left[\int_0^{\infty} dt\, t^{n-1}e^{-t}\right]\left[-\sum_{n=1}^{\infty} {(-z)^n \over n^{\alpha}}\right]=\Gamma(\alpha) F_{\alpha}(z),
\end{eqnarray}
where $\Gamma(\alpha)=\int_0^{\infty} dt\, t^{\alpha-1}e^{-t}$ is the gamma function and $F_{\alpha}(z)=-\sum_{n=1}^{\infty} {(-z)^n \over n^{\alpha}}$. Note that $\Gamma(\alpha)=(\alpha-1)\Gamma(\alpha-1)$ with $\Gamma(1)=1$ and $\Gamma(1/2)=\sqrt{\pi}$, and $F_{\alpha}(z)=z{\partial \over\partial z}F_{\alpha+1}(z)$.

Using the above result, we obtain
\begin{equation} \label{eq:DensityInvar_DoubleBand}
\Delta n=C_\alpha (k_{\rm B} T)^{\alpha} \Gamma(\alpha) \left[F_{\alpha}(z)-F_{\alpha}(z^{-1})\right]
=\frac{C_\alpha}{\alpha}\varepsilon_{\rm F}^\alpha,
\end{equation}
where $z=e^{\beta\mu}$, which is called the fugacity. Thus, finally we have
\begin{equation}
F_{\alpha}(z)-F_{\alpha}(z^{-1}) = \frac{(\beta \varepsilon_{\rm F})^\alpha}{\Gamma(\alpha+1)}.
\end{equation}
By solving the above equation with respect to $z$ for a given $T$, we can obtain the chemical potential $\mu=k_{\rm B} T \ln z$.

At low temperatures, $\beta \mu \rightarrow \infty$ thus $z\rightarrow \infty$. Note that from the Sommerfeld expansion \cite{Ashcroft1976_SM}
\begin{equation} \label{eq:Sommerfeld_expansion}
\lim_{z\rightarrow \infty} \int_0^{\infty}{dx}\frac{H(x)}{z^{-1}e^x+1} \approx \int_0^{\beta \mu}{dx}H(x) + \frac{\pi^2}{6} \frac{\partial H(\beta \mu)}{\partial x} ,
\end{equation}
where $H(x)$ is a function which diverges no more rapidly than a polynomial as $x\rightarrow \infty$.
Then for $H(x) = x^{\alpha-1}$ and using Eq. (\ref{eq:fermionic_integral}), Eq. (\ref{eq:Sommerfeld_expansion}) becomes
\begin{equation}
\lim_{z\rightarrow \infty} F_{\alpha}(z)\approx {(\beta\mu)^{\alpha}\over \Gamma(\alpha+1)}\left[1+{\pi^2\over 6}{\alpha(\alpha-1)\over(\beta\mu)^2}\right],
\end{equation}
whereas $F_{\alpha}(z^{-1})=z^{-1}-{z^{-2}\over 2^{\alpha}}+\cdots$ vanishes as $z\rightarrow \infty$.
Thus, we can obtain the low-temperature correction as
\begin{equation}\label{eq:mu_low_T}
\frac{\mu}{\varepsilon_{\rm F}} \approx 1- \frac{\pi^2}{6} \left(\alpha-1\right) \left(\frac{T}{T_{\rm F}} \right)^2,
\end{equation}
where $T_{\rm F}=\varepsilon_{\rm F}/k_{\rm B}$ is the Fermi temperature.

At high temperatures, $\beta \mu \rightarrow 0$ due to the finite carrier densities, thus $z\rightarrow 1$. From $z\approx 1+\beta\mu+{1\over 2}(\beta\mu)^2$ for $|\beta\mu|\ll 1$,
\begin{equation}
\lim_{z\rightarrow 1} F_\alpha(z)\approx \eta(\alpha)+\eta(\alpha-1)\beta\mu + \frac{1}{2}\eta(\alpha-2)\left(\beta\mu\right)^2,
\end{equation}
where $\eta(\alpha)=F_{\alpha}(1)$ is the Dirichlet eta function \cite{Arfken_SM}. 
Thus, we have $F_{\alpha}(z) - F_{\alpha}(z^{-1}) \approx 2\eta(\alpha-1)\beta\mu$, and obtain the following high-temperature asymptotic form:
\begin{equation}\label{eq:mu_high_T}
\frac{\mu}{\varepsilon_{\rm F}} \approx \frac{1}{2\eta(\alpha-1)\Gamma(\alpha+1)}\left(\frac{T_{\rm F}}{T}\right)^{\alpha-1} .
\end{equation}

For m-WSMs, $\alpha ={2\over J}+1$ and we obtain
\begin{eqnarray}\label{eq:mu_mWSM_temperature_correction}
{\frac{\mu}{\varepsilon_{\rm F}}} &=&
\begin{cases}
1- \frac{\pi^2}{3J} \left(\frac{T}{T_{\rm F}} \right)^2 & (T\ll T_{\rm F}), \\ {1\over 2\eta\left({2\over J}\right)\Gamma\left(2+{2\over J}\right)}\left(\frac{T}{T_{\rm F}}\right)^{-{2\over J}} & (T\gg T_{\rm F}).
\end{cases}
\end{eqnarray}

Next, consider the temperature dependent Thomas-Fermi wavevector $q_{\rm TF}(T)$. Note that in 3D, $q_{\rm TF}^2(0)={4\pi e^2\over \kappa} D(\varepsilon_{\rm F})$ and at finite $T$, $q_{\rm TF}^2(T)={4\pi e^2\over \kappa} {\partial n \over \partial \mu}$. Thus we have
\begin{equation}
\frac{q_{\rm{TF}}^2(T) }{q_{\rm{TF}}^2(0) } = \frac{\partial \varepsilon_{\rm F}}{\partial \mu} =  \frac{\Gamma(\alpha)}{(\beta \varepsilon_{\rm F})^{\alpha-1}}\left[F_{\alpha-1}(z) +F_{\alpha-1}(z^{-1})\right].
\end{equation}
For a given $T$, the chemical potential (or equivalently fugacity $z$) is calculated using the density invariance in Eq.~(\ref{eq:DensityInvar_DoubleBand}), and then $q_{\rm{TF}}(T)$ is obtained from the above relation.

At low temperatures, $\mu(T)$ is given by Eq.~(\ref{eq:mu_low_T}), thus
\begin{eqnarray}\label{eq:q_TF_temperature_correction}
\frac{q_{\rm{TF}}^2(T) }{q_{\rm{TF}}^2(0) }&\approx& \frac{\Gamma(\alpha)}{\left(\beta \varepsilon_{\rm F} \right)^{\alpha-1}} \left[\frac{\left(\beta \mu \right)^{\alpha-1}}{\Gamma(\alpha)} \left(1+\frac{\pi^2}{6}\frac{(\alpha-1)(\alpha-2)}{\left(\beta \mu \right)^{2}} \right) + \cancelto{0}{\left(z^{-1}-\frac{z^{-2}}{2^{\alpha-1}}\right)} \right] \nonumber \\
&\approx& \frac{\mu}{\varepsilon_{\rm F}}+\frac{\pi^2}{6}\frac{(\alpha-1)(\alpha-2)}{\left(\beta\varepsilon_{\rm F} \right)^{2}}
\approx 1-{\pi^2\over 6}(\alpha-1)\left(\frac{T}{T_{\rm F}} \right)^2.
\end{eqnarray}
At high temperatures, $\mu(T)$ is given by Eq.~(\ref{eq:mu_high_T}), thus
\begin{eqnarray}
\frac{q_{\rm{TF}}^2(T) }{q_{\rm{TF}}^2(0) }&\approx&\frac{\Gamma(\alpha)}{\left(\beta \varepsilon_{\rm F} \right)^{\alpha-1}} \left[2\eta(\alpha-1)+\eta(\alpha-3)\left(\beta \mu \right)^{2} \right] \nonumber \\
&\approx& 2\eta(\alpha-1) \Gamma(\alpha)\left(\frac{T}{T_{\rm F}} \right)^{\alpha-1}.
\end{eqnarray}

For m-WSMs, we find
\begin{eqnarray}\label{eq:q_TF_mWSM_temperature_correction}
{\frac{q_{\rm{TF}}(T) }{q_{\rm{TF}}(0) }} &= &
\begin{cases}
1- \frac{\pi^2}{6J} \left(\frac{T}{T_{\rm F}} \right)^2  & (T\ll T_{\rm F}), \\
\sqrt{2\eta\left({2\over J}\right)\Gamma\left(1+{2\over J}\right)}\left(\frac{T}{T_{\rm F}}\right)^{1\over J} & (T\gg T_{\rm F}),
\end{cases}
\end{eqnarray}
where $q_{\rm TF}(0)=q_{\rm TF}$ is given by Eq.~(\ref{eq:q_TF_mWSM}).

\begin{figure}[htb]
\includegraphics[width=0.6\linewidth]{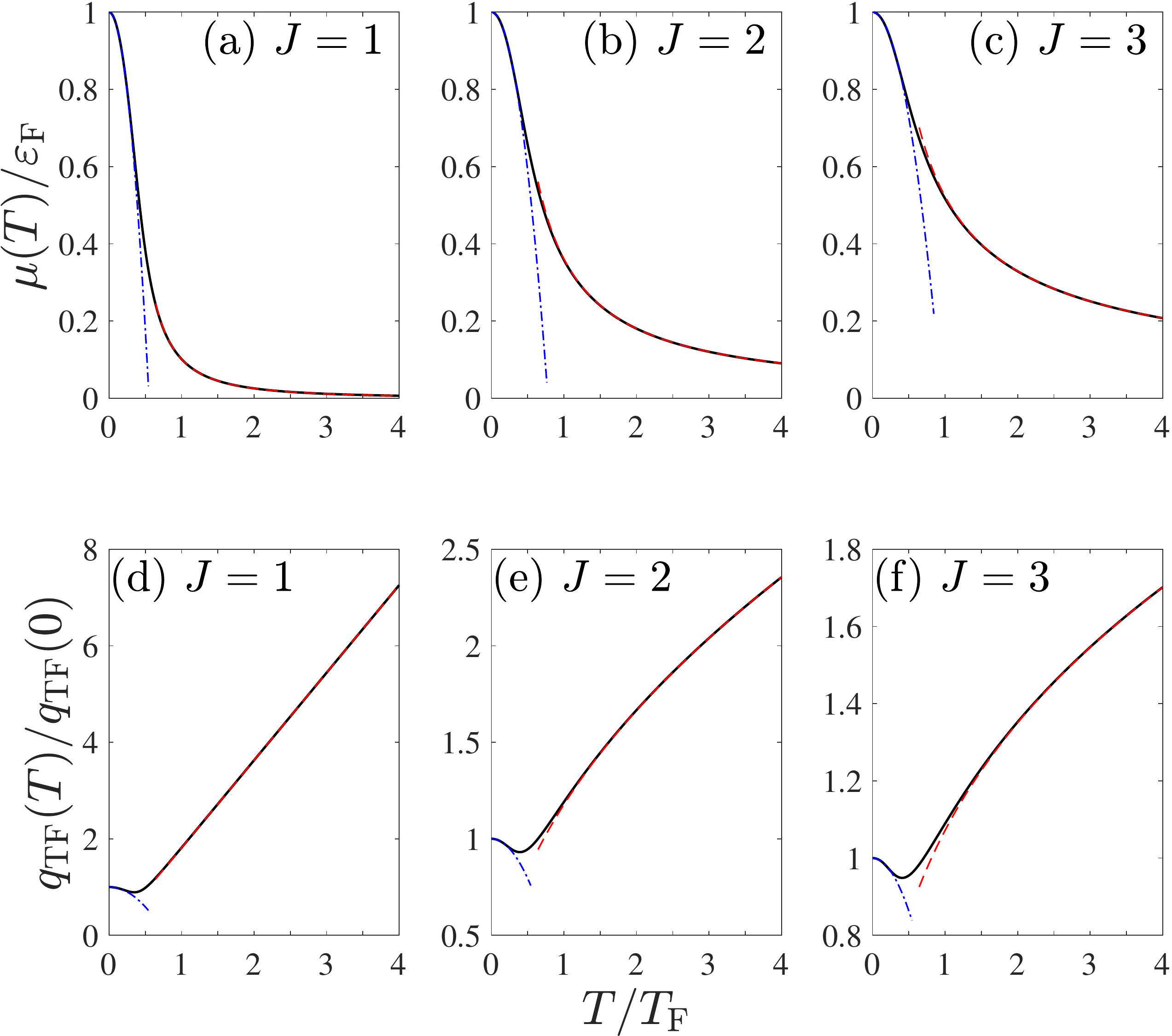}
\caption{
Temperature dependence of (a)-(c) chemical potential and (d)-(f) Thomas-Fermi wavevector for m-WSMs with $J=1,2,3$. Red dashed and blue dashed-dotted lines represent the asymptotic forms in
Eqs.~(\ref{eq:mu_mWSM_temperature_correction}) and (\ref{eq:q_TF_mWSM_temperature_correction}).
}
\label{fig:temperature_dependence_thomas_fermi_wavevector}
\end{figure}

Figure \ref{fig:temperature_dependence_thomas_fermi_wavevector} shows the temperature dependence of the chemical potential and Thomas-Fermi wavevector in m-WSMs.

\section{Temperature dependence of dc conductivity in multi-Weyl semimetals}
\label{sec:temperature_dependence}

From Eq.~(\ref{eq:conductivity_tensor}) in the main text, we can easily generalize the conductivity tensor at zero temperature to that at finite temperature. For $f^{(0)}(\varepsilon)=\left[z^{-1}e^{\beta \varepsilon}+ 1\right]^{-1}$, $S^{(0)}(\varepsilon)=-{\partial f^{(0)}(\varepsilon)\over\partial\varepsilon}=\beta f^{(0)}(\varepsilon) \left(1-f^{(0)}(\varepsilon)\right)={\beta z^{-1} e^{\beta \varepsilon} \over (z^{-1} e^{\beta\varepsilon}+ 1)^2}$.
Then the conductivity tensor at finite temperature is given by
\begin{eqnarray}
\label{eq:conductivity_finite_temperature}
\sigma_{ij}(T)&=&g e^2\int {d^3 k\over (2\pi)^3} \left(-{\partial f^{(0)}(\varepsilon_{\bm{k}})\over\partial\varepsilon}\right) v_{\bm{k}}^{(i)} v_{\bm{k}}^{(j)}\tau_{\bm{k}}^{(j)} \nonumber \\
&=&g e^2 \int_0^{\infty}dr \int_0^{\pi}d\theta \int_0^{2\pi}d\phi {k_0^3 r^{2\over J}\sin^{{2\over J}-1}\theta\over (2\pi)^3 c_z J}
{\beta z^{-1} e^{\beta \varepsilon_0 r} \over (z^{-1} e^{\beta\varepsilon_0 r}+ 1)^2} v_{\bm{k}}^{(i)} v_{\bm{k}}^{(j)}\tau_{\bm{k}}^{(j)} \nonumber \\
&=&{\sigma_0 \over J} \int_0^{\infty}dr r^{2\over J} \int_{-1}^{1}d\mu \left(1-\mu^2\right)^{{1\over J}-1} \int_0^{2\pi}{d\phi\over 2\pi} {\beta\varepsilon_0 z^{-1} e^{\beta \varepsilon_0 r} \over (z^{-1} e^{\beta\varepsilon_0 r}+ 1)^2} \tilde{v}_{\bm{k}}^{(i)} \tilde{v}_{\bm{k}}^{(j)} \tilde{\tau}_{\bm{k}}^{(j)}.
\end{eqnarray}
Thus from Eq.~(\ref{eq:band_velocity}), we have
\begin{subequations}
\begin{eqnarray}
\sigma_{xx}(T)&=&\sigma_0 {J\over 2} \int_0^{\infty} dr\, r^2 {\beta\varepsilon_0 z^{-1} e^{\beta \varepsilon_0 r} \over (z^{-1} e^{\beta\varepsilon_0 r}+ 1)^2} \int_{-1}^{1}d\mu \left(1-\mu^2\right)\tilde{\tau}^{(x)}(\mu), \\
\sigma_{zz}(T)&=& \sigma_0 {c_z^2 \over J} \int_0^{\infty} dr\, r^{2\over J}{\beta\varepsilon_0 z^{-1} e^{\beta \varepsilon_0 r} \over (z^{-1} e^{\beta\varepsilon_0 r}+ 1)^2} \int_{-1}^{1}d\mu \left(1-\mu^2\right)^{{1\over J}-1}\mu^2 \tilde{\tau}^{(z)}(\mu).
\end{eqnarray}
\end{subequations}

To derive the asymptotic behaviors of $\sigma_{ii}(T)/\sigma_{ii}(0)$ at low and high temperatures, let us rewrite Eq.~(\ref{eq:conductivity_tensor}) in the main text, in the following energy integral form:
\begin{eqnarray}
\label{eq:conductivity_finite_temperature_energy_int}
\sigma_{ii}(T)&=&g e^2 I \int_0^{\infty} d\varepsilon \left(-{\partial f^{(0)}(\varepsilon)\over\partial\varepsilon}\right) D(\varepsilon) [v^{(i)}(\varepsilon)]^2 \tau^{(i)}(\varepsilon,T),
\end{eqnarray}
where $I$ is a factor from the angular integration. Note that the factor $I$ will be canceled by $\sigma_{ii}(0)$ later. Assuming that $\tau^{(i)}(\varepsilon, T)$ can be decomposed as
\begin{equation}\label{eq:tau_decomposition}
\tau^{(i)}(\varepsilon, T) = \tau^{(i)}(\varepsilon) g^{(i)} \left(\frac{T}{T_{\rm F}}\right),
\end{equation}
where $g^{(i)} \left(\frac{T}{T_{\rm F}}\right)$ is the energy-independent correction term from the screening effect with $g^{(i)}(0)\equiv1$, we can separate the contributions from the energy averaging over the Fermi distribution and the temperature dependent screening. Suppose $D(\varepsilon) \propto \varepsilon^{\alpha-1}$, $v^{(i)}(\varepsilon) \propto \varepsilon^{\nu}$, and $ \tau^{(i)}(\varepsilon) \propto \varepsilon^{\gamma}$. Then we can express $\sigma_{ii}(T)$ as
\begin{eqnarray}
\label{eq:conductivity_finite_temperature_energy_int_2}
\sigma_{ii}(T)&=&C \int_0^{\infty} d\varepsilon \left(-{\partial f^{(0)}(\varepsilon)\over\partial\varepsilon}\right) \varepsilon^{\alpha-1+2\nu+\gamma} g^{(i)} \left(\frac{T}{T_{\rm F}}\right) \nonumber \\
&=& C (k_{\rm B}T)^{\delta}\Gamma(\delta+1)F_{\delta}(z)g \left(\frac{T}{T_{\rm F}}\right),
\end{eqnarray}
where $C$ is a constant and $\delta \equiv \alpha-1+2\nu+\gamma$. Note that Eq.~(\ref{eq:conductivity_finite_temperature_energy_int_2}) reduces to $\sigma_{ii}(0)= C \varepsilon_{\rm F}^{\delta}$ at zero temperature.
Therefore, after eliminating $C$, we have
\begin{equation}
{\sigma_{ii}(T)\over \sigma_{ii}(0)}= \frac{\Gamma(\delta+1)F_{\delta}(z)}{(\beta\varepsilon_{\rm F})^{\delta}} g^{(i)} \left(\frac{T}{T_{\rm F}}\right).
\end{equation}
For short-range impurities, $g^{(i)}\left({T\over T_{\rm F}}\right)=1$.
For charged impurities at low temperatures, from the form of the low-temperature correction for the Thomas-Fermi wavevector in Eq.~(\ref{eq:q_TF_temperature_correction}), we expect
\begin{equation}
g^{(i)}\left({T\over T_{\rm F}}\right)\approx 1-A^{(i)}\left({T\over T_{\rm F}}\right)^2.
\end{equation}
Note that $A^{(i)}$ depends on the screening strength, and in the strong screening limit, from Eq.~(\ref{eq:q_TF_mWSM_temperature_correction}) we have $A^{(i)}={2\pi^2\over 3J}$.
At high temperatures, however, $\tau^{(i)}(\varepsilon,T)$ cannot be simply decomposed as Eq.~(\ref{eq:tau_decomposition}). The energy averaging typically dominates over the screening contribution \cite{DasSarma2015_SM}, and the screening correction $g\left({T\over T_{\rm F}}\right)$ only gives a constant factor without changing the temperature power. Assuming $g^{(i)}\left({T\over T_{\rm F}}\right)\approx 1$ at high temperatures, then in the low and high temperature limits, we have
\begin{eqnarray}\label{eq:asymptotic_temp}
{\sigma_{ii}(T)\over \sigma_{ii}(0)} &=&
\begin{cases}
1+\left[\frac{\pi^2}{6}(\delta-\alpha)\delta-A^{(i)}\right] \left(\frac{T}{T_{\rm F}}\right)^2  & (T\ll T_{\rm F}), \\
\Gamma(\delta+1) \eta(\delta)\left(\frac{T}{T_{\rm F}}\right)^{\delta}  & (T\gg T_{\rm F}).
\end{cases}
\end{eqnarray}

\begin{figure}[htb]
\includegraphics[width=0.6\linewidth]{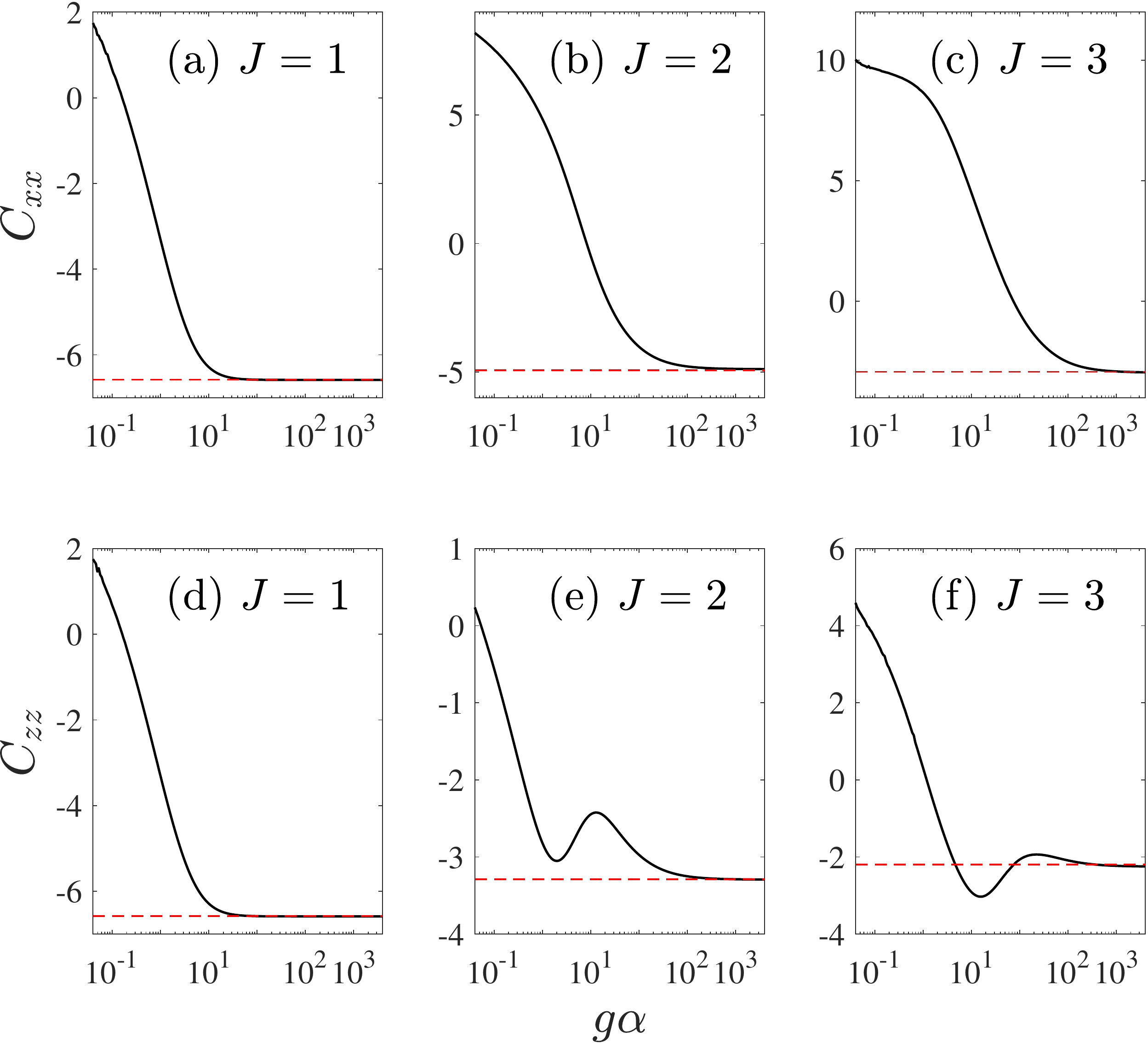}
\caption{
Low-temperature coefficients (a)-(c) $C_{xx}$ and (d)-(f) $C_{zz}$ as a function of the screening strength $g\alpha$ for charged impurities. Red dashed lines represent the low-temperature coefficients in the strong screening limit given by Eq.~(\ref{eq:temperature_dependence_charged_strong_screening}). Here, $n=n_0$ is used for calculation.
}
\label{fig:temperature_coefficient_evolution}
\end{figure}

Now, consider m-WSM with $\alpha=\frac{2}{J}+1$.
For short-range impurities, $g^{(i)} \left(\frac{T}{T_{\rm F}} \right)=1$, and from the energy dependence of the relaxation time in Eq.~(\ref{eq:tau_short_analytic}), $\gamma=-\frac{2}{J}$. Thus, we find
\begin{subequations}\label{eq:temperature_dependence_short_asym}
\begin{eqnarray}
{\sigma_{xx}(T)\over \sigma_{xx}(0)} &=&
\begin{cases}
1+ \frac{\pi^2}{3}\left(\frac{J-1}{J}\right)\left(\frac{J-4}{J}\right)\left({T\over T_{\rm F}}\right)^2  & (T\ll T_{\rm F}), \\
\Gamma(3-\frac{2}{J})\eta(2-\frac{2}{J})\left({T\over T_{\rm F}}\right)^{2-\frac{2}{J}} & (T\gg T_{\rm F}),
\end{cases} \\
{\sigma_{zz}(T)\over \sigma_{zz}(0)} &=&
\begin{cases}
1-e^{-T_{\rm F}/T}  & (T\ll T_{\rm F}), \\
\frac{1}{2}+ \frac{1}{8\eta\left({2\over J}\right)\Gamma\left({2\over J}+2\right)} \left(\frac{T}{T_{\rm F}}\right)^{-{2+J\over J}} & (T\gg T_{\rm F}).
\end{cases}\end{eqnarray}
\end{subequations}

For charged impurities in the strong screening limit, from Eq.~(\ref{eq:tau_strong_screening}), $\tau^{(i)}(\varepsilon)\sim \varepsilon^{-{2\over J}}$ thus $\gamma=-\frac{2}{J}$ at low temperatures, whereas at high temperatures $\tau^{(i)}(\varepsilon)\sim \varepsilon^{-{2\over J}+{4\over J}}$ because thermally induced charge carriers participate in transport giving $\gamma=\frac{2}{J}$. Combining the temperature dependent screening correction with $A^{(i)}={2\pi^2\over 3J}$ at low temperatures, we find
\begin{subequations}\label{eq:temperature_dependence_charged_strong_screening}
\begin{eqnarray}
{\sigma_{xx}(T)\over \sigma_{xx}(0)} &=&
\begin{cases}
1+ \frac{\pi^2}{3}\left(\frac{J^2-7J+4}{J^2}\right)\left({T\over T_{\rm F}}\right)^2   & (T\ll T_{\rm F}), \\
\Gamma(3+\frac{2}{J})\eta(2+\frac{2}{J})\left({T\over T_{\rm F}}\right)^{2+\frac{2}{J}} & (T\gg T_{\rm F}),
\end{cases} \\
{\sigma_{zz}(T)\over \sigma_{zz}(0)} &=&
\begin{cases}
1-\frac{2\pi^2}{3J}\left({T\over T_{\rm F}}\right)^2 & (T\ll T_{\rm F}), \\
\Gamma(1+\frac{4}{J})\eta(\frac{4}{J})\left({T\over T_{\rm F}}\right)^{\frac{4}{J}} & (T\gg T_{\rm F}).
\end{cases}\end{eqnarray}
\end{subequations}

For charged impurities at arbitrary screening, from the Fermi energy dependence of the relaxation time discussed in Sec.~\ref{sec:conductivity_3D_anisotropic}, $\gamma=4\zeta-\frac{2}{J}$ with ${1\over J} \le \zeta \le 1$ at high temperatures. Thus, we can express the low and high temperature asymptotic forms as
\begin{subequations}\label{eq:temperature_dependence_charged_arbitrary_screening}
\begin{eqnarray}
{\sigma_{xx}(T)\over \sigma_{xx}(0)} &=&
\begin{cases}
1+ C_{xx}\left({T\over T_{\rm F}}\right)^2 & (T\ll T_{\rm F}), \\
\Gamma\left(3+4\zeta-{2\over J}\right)\zeta\left(2+4\zeta-{2\over J}\right)\left({T\over T_{\rm F}}\right)^{2+4\zeta-{2\over J}} & (T\gg T_{\rm F}),
\end{cases} \\
{\sigma_{zz}(T)\over \sigma_{zz}(0)} &=&
\begin{cases}
1+ C_{zz}\left({T\over T_{\rm F}}\right)^2 & (T\ll T_{\rm F}), \\
\Gamma(1+4\zeta)\eta(4\zeta)\left({T\over T_{\rm F}}\right)^{4\zeta} & (T\gg T_{\rm F}).
\end{cases}\end{eqnarray}
\end{subequations}
As explained in Sec.~\ref{sec:conductivity_3D_anisotropic}, $\zeta$s in $\sigma_{xx}$ and $\sigma_{zz}$ do not need to be the same.
Note that $\zeta={1\over J}$ in Eq.~(\ref{eq:temperature_dependence_charged_arbitrary_screening}) gives the same high-temperature exponent corresponding to the strong screening limit in Eq.~(\ref{eq:temperature_dependence_charged_strong_screening}), and the temperature dependent conductivity has the high-temperature asymptotic form given by Eq.~(\ref{eq:temperature_dependence_charged_arbitrary_screening}) with $\zeta$ which varies within ${1\over J}\le \zeta \le 1$ and approaches ${1\over J}$ in the strong screening limit.

Figure \ref{fig:temperature_coefficient_evolution} shows the evolution of the low-temperature coefficients $C_{xx}$ and $C_{zz}$ in Eq.~(\ref{eq:temperature_dependence_charged_arbitrary_screening}) for charged impurities as a function of the screening strength $g\alpha$.
Above a critical $g\alpha$, $C_{xx}$ and $C_{zz}$ become negative, thus the conductivity decreases with temperature, showing a metallic behavior. As $g\alpha$ increases further, the low-temperature coefficients eventually approach $C_{xx}=\frac{\pi^2}{3}\left(\frac{J^2-7J+4}{J^2}\right)$ and $C_{zz}=-\frac{2\pi^2}{3J}$, as obtained in Eq.~(\ref{eq:temperature_dependence_charged_strong_screening}). The non-monotonic behavior in the low-temperature coefficients $C_{zz}$ as a function of $g\alpha$ for $J>1$ originates from the angle-dependent power-law in the relaxation time, similarly as shown in Fig.~\ref{fig:density_power_evolution} in the main text.


\end{document}